\documentclass[journal]{IEEEtran}
\usepackage{amsfonts}
\usepackage{bm}
\usepackage{amsmath,amsfonts,amsthm,amssymb}
\usepackage{setspace}
\usepackage{Tabbing}
\usepackage{fancyhdr}
\usepackage{lastpage}
\usepackage{extramarks}
\usepackage{chngpage}
\usepackage{soul,color}
\usepackage{epsfig}
\usepackage{graphicx,float,wrapfig,subfigure}
\usepackage{dsfont}
\usepackage{longtable}
\usepackage{wrapfig}
\usepackage{stfloats}
\usepackage{cite}
\usepackage{graphicx}
\usepackage{array}
\usepackage{multirow}
\usepackage{multicol}
\usepackage{colortbl}
\usepackage{tabularx}
\usepackage{mdwmath}
\usepackage{mdwtab}
\usepackage{color}
\usepackage{verbatim}
\usepackage{amsmath}
\usepackage{tikz}
\usetikzlibrary{calc}
\usepackage{flowchart}
\usetikzlibrary{shapes.geometric, arrows}
\usepackage{overpic}
\usepackage{epstopdf}
\usepackage{url}
\usepackage[colorlinks,linkcolor=black,anchorcolor=black,citecolor=black,urlcolor=black]{hyperref} 

\usepackage{makecell}
\usepackage{textcomp}
\usepackage{bbm}

\usepackage{enumerate}
\usepackage[ruled,vlined,linesnumbered]{algorithm2e} 

\newtheorem{remark}{Remark}

\usepackage{tikz}
\usetikzlibrary{patterns, arrows.meta, decorations.pathreplacing}

\usepackage{threeparttable}

\usepackage{mathtools}

\usepackage{color}
\makeatletter %
\let\myorg@bibitem\bibitem
\def\bibitem#1#2\par{%
	\@ifundefined{bibitem@#1}{%
		\myorg@bibitem{#1}#2\par
	}{%
		\begingroup
		\color{\csname bibitem@#1\endcsname}%
		\myorg@bibitem{#1}#2\par
		\endgroup
	}%
}

\makeatother %

\allowdisplaybreaks

\usepackage{footnote}
\makesavenoteenv{tabular}
\makesavenoteenv{table}
\usepackage{booktabs}

\begin{document}

\title{
A Convexified Eulerian Framework for Scalable Coordination of Massive DER Populations
	}

\author{
Ge~Chen,~\IEEEmembership{Member,~IEEE,}
Yiwei~Qiu,~\IEEEmembership{Member,~IEEE,}
Shiyao~Zhang,~\IEEEmembership{Senior Member,~IEEE,}
Pengfei~Su,~\IEEEmembership{ Member,~IEEE,}
Haoran~Deng,~\IEEEmembership{ Member,~IEEE,}
and~Hongcai~Zhang,~\IEEEmembership{Senior Member,~IEEE}
 \vspace{-6mm}
\thanks{
%
G. Chen, S. Zhang, P. Su, and H. Deng are with the School of Advanced Engineering, Great Bay University, Dongguan, China. Y. Qiu is with the
College of Electrical Engineering, Sichuan University, Chengdu 610065, China.
H. Zhang is with the State Key Laboratory of Internet of Things for Smart City, University of Macau, Macao, 999078 China. 
}
}

\maketitle

\begin{abstract}
	This paper proposes a scalable coordination framework with aggregator-side privacy protection for storage-like distributed energy resources (DERs). The framework adopts a two-layer architecture. At the macroscopic layer, building upon an \emph{Eulerian} modeling perspective, the DER population is represented as a continuum whose density evolution is governed by a partial differential equation (PDE), such that the computational complexity is independent of the population size. To address the bilinear non-convexity in this PDE-constrained optimization problem, we develop a convexification method that combines finite-volume discretization with a \emph{flux-lifting} technique, reformulating the macroscopic problem into a sparse linear program (LP). The LP solution yields a unified, state-dependent broadcast signal for population coordination. Furthermore, a Wasserstein-based relaxation is introduced to replace rigid cyclic constraints and provide additional operational flexibility for improved economic performance. At the microscopic layer, individual resources autonomously recover local setpoints from the broadcast signal and their local states, while an upstream data-mixing protocol aggregates individual states into a macroscopic density histogram without exposing raw individual states to the aggregator. Numerical studies validate the scalability, feasibility, and economic effectiveness of the proposed framework.
\end{abstract}

\begin{IEEEkeywords}
	Distributed energy resources, Eulerian perspective, finite-volume method, flexibility aggregation, partial differential equations.
\end{IEEEkeywords}

\section{Introduction}
\label{sec:intro}

The global imperative to mitigate climate change has accelerated the transition toward low-carbon energy systems, marked by high renewable penetration and widespread electrification of end-use sectors \cite{powell2022charging}. Because renewable generation is inherently intermittent, distributed energy resources (DERs) have become an important source of grid flexibility \cite{chen2025quantifying}. Among them, \emph{storage-like} resources, such as electric vehicles (EVs), battery energy storage systems (BESSs), and thermostatically controlled loads (TCLs), are particularly attractive because they can buffer energy and shift demand over time \cite{chen2025quantifying}. By coordinating such DER populations, aggregators can facilitate renewable integration \cite{10058886}, support peak-load shifting \cite{li2024impact}, and reduce electricity costs \cite{9800974}.

\begin{figure}[H]
	\vspace{-4mm}
	\centering	{\includegraphics[width=1\columnwidth]{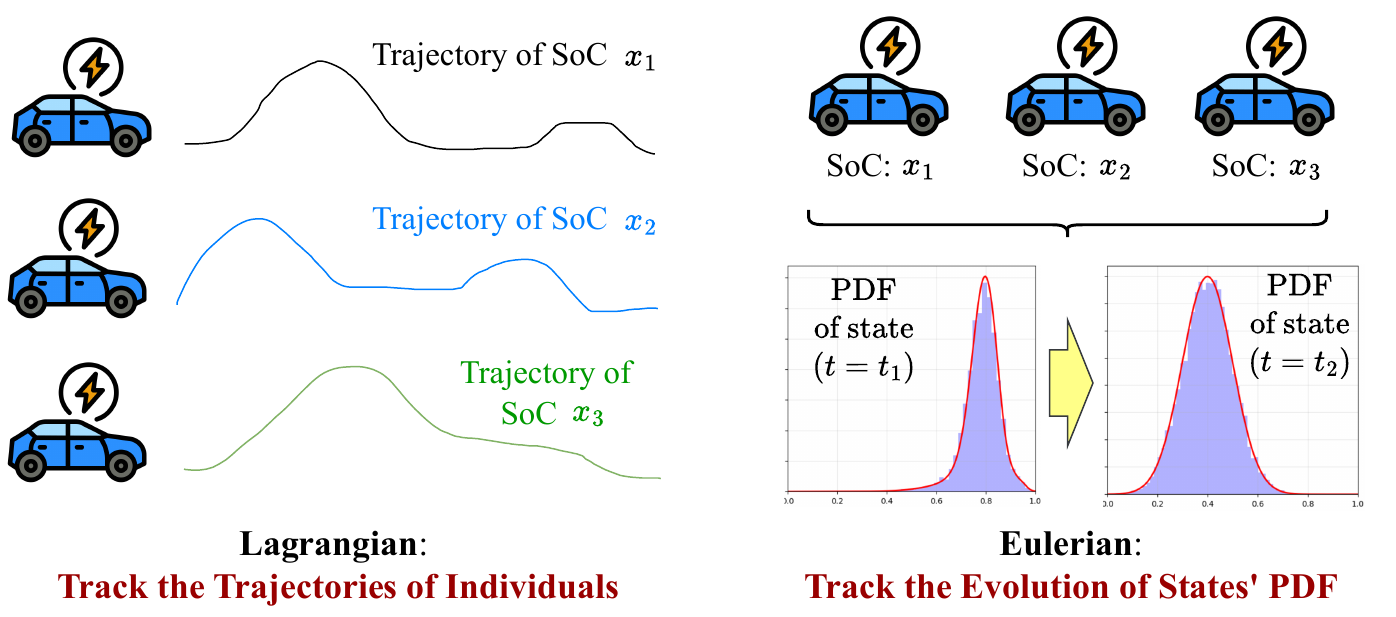}}
	\vspace{-8mm}
	\caption{Schematic comparison between the Lagrangian and Eulerian perspectives. The Lagrangian approach (left) tracks the microscopic state trajectories of individual resources, whereas the Eulerian approach (right) characterizes the macroscopic evolution of the population's state PDF.}
	\vspace{-2mm}
	\label{fig_MF}
\end{figure}

Existing methodologies for coordinating DER populations can be broadly grouped into two paradigms \cite{9914677}: the \emph{Lagrangian} perspective and the \emph{Eulerian} perspective. Their main difference lies in modeling granularity, as illustrated in Fig.~\ref{fig_MF}. The Lagrangian perspective is \emph{device-centric}: it treats the population as a collection of distinct agents and tracks the state trajectories (e.g., the SoC of EVs) of individual devices \cite{11023621,11004617}. In contrast, the Eulerian perspective is \emph{population-centric}: it treats the population as a continuum and tracks the evolution of the probability density function (PDF) of the states across the entire population \cite{10741829}. Borrowing an analogy from transportation systems, the Lagrangian approach is akin to tracking the GPS trajectory of every vehicle, whereas the Eulerian approach monitors traffic density on road segments without identifying individual cars.

The \emph{Lagrangian} perspective has long been the dominant framework in existing literature \cite{li2026novel}. It formulates a joint optimization problem whose decision variables are the power setpoints of all devices over the entire horizon. This approach can readily accommodate detailed physical constraints and resource heterogeneity. For instance, EV fleets have been coordinated by explicitly controlling the charging behavior of individual vehicles \cite{10614321}, and similar individual-level formulations have been used for TCL populations \cite{11106947}. However, it suffers from the ``curse of dimensionality'': the problem size scales linearly with the population size $N$ \cite{9766186}. As $N$ grows to millions, the resulting optimization becomes computationally intractable \cite{9975832}. Moreover, effective Lagrangian coordination requires users to share their real-time states with the aggregator \cite{9760472}. This exposure raises privacy concerns about user behavior patterns \cite{10580944, 10251779}.

To overcome these scalability and privacy bottlenecks, a growing body of work has turned to \emph{Eulerian} macroscopic modeling frameworks \cite{10741829}. Rather than optimizing the trajectories of individual devices, this paradigm approximates a large population as a continuum and uses partial differential equations (PDEs), such as the Fokker-Planck equation (FPE), to describe the spatiotemporal advection and diffusion of the population state density \cite{10909456}. Early works used such PDE models for aggregate load management \cite{6239581, perfumo2012load}, and more recent studies have extended this viewpoint to reinforcement learning \cite{10851340} and mean-field games \cite{11029625, 8515099}. Its main advantage is computational scalability: the resulting optimization size does not depend on the population size $N$ \cite{caballe2026large}. In addition, the aggregator coordinates the population through a broadcast control signal rather than individualized commands, which can reduce uplink state exposure by avoiding direct access to raw individual states \cite{10740528}.

Despite these modeling advantages, existing Eulerian frameworks still face three critical bottlenecks in multi-period grid-facing dispatch. First, the governing PDEs typically contain \emph{bilinear} advective terms, i.e., products of control decisions and probability-density variables \cite{6239581}. While such terms can be handled in nonlinear tracking settings, they make the resulting scheduling problem inherently non-convex and difficult to solve for complex economic dispatch \cite{8424013}. Second, standard approaches often rely on unified, \emph{state-independent} broadcast signals to preserve privacy \cite{perfumo2012load}. Without \emph{state-dependency}, resources cannot differentiate their responses, which may cause subset actions to cancel each other and thus degrade aggregate flexibility and induce internal energy losses. Third, to ensure long-term operability, conventional studies usually impose \emph{hard pointwise} cyclic constraints that require the terminal probability density to exactly match the initial distribution \cite{9639978}. This rigidity overlooks the economic value of allowing controlled distributional deviations. Therefore, \emph{developing a scalable Eulerian framework that supports state-dependent execution and economically meaningful near-cyclicity remains an open question.}

To bridge these gaps, this paper proposes a two-layer Eulerian coordination framework with aggregator-side privacy protection for large-scale storage-like DER populations. At the macroscopic scheduling layer, we adopt an Eulerian representation to characterize the population density evolution for coordination at the aggregator/point of common coupling (PCC) interface, and formulate the scheduling task as a PDE-constrained optimization problem whose size is independent of the population size. At the microscopic execution layer, we design a decentralized mechanism that maps macroscopic broadcast signals to autonomous individual actions. The main contributions of this paper are threefold:

\begin{enumerate}
	\item We develop a convexification method that projects probability-mass transport from the state space into a flux space via a \emph{flux-lifting} technique, thereby overcoming the inherent bilinear non-convexity of the PDE-constrained aggregator-coordination problem. Combined with finite-volume discretization, this projection reformulates the discretized macroscopic optimization as an efficient sparse linear program (LP) while preserving physical consistency and probability-mass balance.
	
	\item We design a decentralized execution mechanism driven by macroscopic \emph{state-dependent} broadcast signals. Guided by this signal, individual resources use local inverse-dynamics mappings to autonomously compute distinct power setpoints. A data-mixing protocol is further integrated, so individual states are aggregated into an anonymous distribution before reaching the aggregator. This enables population flexibility to be used without requiring the aggregator to access raw individual states.
	
	\item We propose a near-cyclicity mechanism that uses the Wasserstein metric to quantify and bound macroscopic distributional deviations, thereby unlocking the economic flexibility typically suppressed by strict pointwise cyclic constraints. This formulation gives the aggregator a tunable tool to trade off terminal energy offsets against operating-cost reductions.
\end{enumerate}

The remainder of this paper is organized as follows. Section~\ref{sec:overview} provides the method overview. Section~\ref{sec:modeling} and Section~\ref{sec:solver} present the Eulerian model and LP reformulation, respectively. Section~\ref{sec:dispatch} delineates the decentralized execution. Numerical results are given in Section~\ref{sec:case}, and conclusions are drawn in Section~\ref{sec:conclusion}.

\section{Method Overview}
\label{sec:overview}

This paper develops a scalable coordination framework with aggregator-side privacy protection for large-scale \emph{storage-like} DER populations. The key idea is to optimize the \emph{macroscopic} density evolution of the population through an Eulerian model while enabling \emph{microscopic} decentralized execution without tracking individual devices. This separation of roles is central to the framework: the aggregator handles a population-level scheduling problem whose size is independent of the number of devices, whereas each device keeps its own state locally and converts a common broadcast signal into an individualized action.\footnote{Although the following method is presented for a \emph{single} population for clarity, the proposed interface can in principle be instantiated for various DER categories by redefining the state variables and parameters. When microscopic heterogeneity is significant, the population can be pre-clustered and each cluster can be modeled as a representative population.} The main notation is summarized in Table~\ref{tab:notation}.

\begin{table}[!t]
	\vspace{-4mm}
	\caption{Nomenclature}
	\label{tab:notation}
	\vspace{-2mm}
	\footnotesize
	\centering
	\renewcommand{\arraystretch}{1.05}
	\begin{tabular}{l l}
		\toprule
		\textbf{Symbol} & \textbf{Description} \\
		\midrule
		\multicolumn{2}{l}{\textit{Indices and Sets}} \\
		$i \in \mathcal{N}$ & Index of agents in the population \\
		$t \in \mathcal{T}$ & Index of discrete time steps \\
		$k \in \mathcal{K}$ & Index of spatial control volumes \\
		\midrule
		\multicolumn{2}{l}{\textit{Parameters}} \\
		$x_i(t)$ & Normalized state of agent $i$ \\
		$f(x)$ & Natural drift dynamics \\
		$\gamma$ & Energy-to-state conversion coefficient \\
		$D$ & Effective diffusion coefficient \\
		$P^{\min/\max}$ & Power rating limits \\
		$E^{\mathrm{total}}$ & Total aggregated capacity scale of the population \\
		$\mathbf{M}$ & Discrete divergence matrix \\
		$\mathbf{L}$ & Discrete diffusion matrix \\
		\midrule
		\multicolumn{2}{l}{\textit{Decision Variables}} \\
		$\rho(x,t)$ & Macroscopic probability density function \\
		$v(x,t)$ & Net drift velocity field (control policy) \\
		$u(x,t)$ & State-dependent active-power policy \\
		$\Phi^{\mathrm{adv}}$ & Advective probability flux (lifted variable) \\
		$P^{\mathrm{agg}}(t)$ & Aggregate active power load of the population \\
		$P^{\mathrm{g}}(t)$ & Net power exchange with the grid \\
		\bottomrule
	\end{tabular}
	\vspace{-4mm}
\end{table}

\begin{figure}[H]
	\centering
	\vspace{-5mm}
	\includegraphics[width=1\columnwidth]{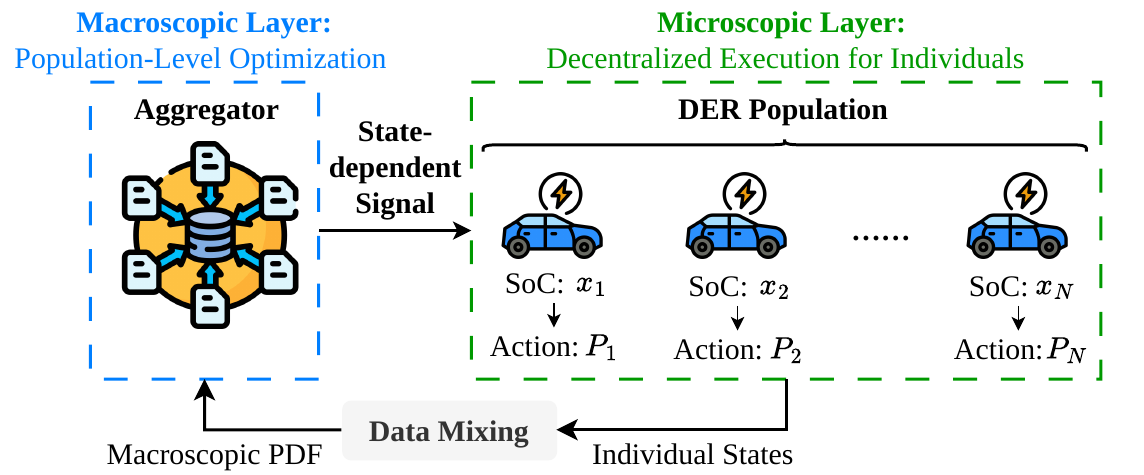}
	\vspace{-7mm}
	\caption{Overview of the proposed two-layer architecture for coordinating a \emph{storage-like} DER population. An EV population is used as an example, but the architecture can be applied to other storage-like DER categories.}
	\label{fig_overview}
	\vspace{-4mm}
\end{figure}

To coordinate a DER population at scale, we adopt a two-layer architecture that decouples the aggregator's optimization from resource-level actuation, as shown in Fig.~\ref{fig_overview}:
\begin{itemize}
	\item \emph{Macroscopic Optimization Layer (Aggregator)}:
	The aggregator adopts the \emph{Eulerian} perspective and represents the DER population as a continuum over a normalized state space. Using the anonymously aggregated initial state distribution, it solves a macroscopic optimization problem that determines the optimal density evolution while minimizing grid-facing operating cost subject to physical constraints. The solution is then converted into a \emph{state-dependent} control signal that guides the collective behavior of the population.
	
	\item \emph{Microscopic Execution Layer (Individual resources)}:
	Upon receiving the broadcast signal, each resource locally computes its own power setpoint from its current state via a unified inverse-dynamics rule. This decentralized execution allows different devices to respond differently without individualized commands.
\end{itemize}

The macroscopic optimizer requires the initial distribution and feedback on the population state evolution. Traditionally, this would require individual resources to share their private state trajectories. To reduce such exposure, we introduce an upstream data-mixing mechanism: individual states are aggregated into an anonymous macroscopic density histogram before reaching the aggregator. In this way, the aggregator can monitor the collective distribution and close the control loop without directly observing the raw states of specific individuals. This design is compatible with state-dependent execution because the aggregator only needs the population-level density, whereas each device reconstructs its own control action locally from the broadcast signal and its current state. At this interface, the information handled by the aggregator scales with the histogram resolution rather than with the number of devices. The privacy claim here should therefore be interpreted as interface-level exposure reduction rather than as a formal cryptographic guarantee.

\section{Macroscopic Optimization}
\label{sec:modeling}

Building upon the classic Eulerian modeling paradigm \cite{6239581, perfumo2012load}, this section formulates the macroscopic optimization model. We first define a unified stochastic model for individual resources. Aggregating these dynamics yields the governing PDEs that characterize the macroscopic density evolution of the population. Based on these physical models, we then construct a PDE-constrained optimization problem whose solution provides the optimal \emph{state-dependent broadcast signal} for decentralized execution.

Before detailing the formulation, we state three assumptions that define the model scope: i) \emph{Representative population}: the macroscopic scheduler is constructed for a homogeneous population, or for a representative population obtained after pre-clustering when microscopic heterogeneity is significant; ii) \emph{Short-term stochasticity}: additive white noise captures high-frequency uncertainties (e.g., user randomness), explicitly excluding long-term degradation or severe anomalies; iii) \emph{Well-posedness}: the control signal is Lipschitz continuous.

\vspace{-2mm}
\subsection{Microscopic Stochastic Dynamics}
Consider an aggregator coordinating a large population of $N$ storage-like resources. Let $i \in \mathcal{N}$ and $t \in \mathcal{T}$ represent the resource and time indices, respectively, and let $x_i(t)\in\mathcal{X}\triangleq[0,1]$ denote the normalized state of the $i$-th agent at time $t$.

\subsubsection{Dynamic Model}
The state evolution of the $i$-th resource is governed by the It\^{o} stochastic differential equation \cite{LU2021268}:
\begin{align}
	\hspace{-2mm}
	\mathrm{d} x_i(t) = v\big(x_i(t),t\big)\,\mathrm{d}t + \sqrt{2D}\,\mathrm{d}W_i(t),
	\ \forall i\in\mathcal{N},\forall t\in\mathcal{T},
	\label{eq:micro_sde}
\end{align}
where the state change $\mathrm{d}x_i(t)$ is decomposed into a deterministic trend $v\big(x_i(t),t\big)\,\mathrm{d}t$ and a stochastic diffusion $\sqrt{2D}\,\mathrm{d}W_i(t)$; $v(\cdot)$ denotes the deterministic change rate, $W_i(t)$ is a standard Wiener process that models resource-level uncertainty, and $D$ is the effective diffusion coefficient that determines the magnitude of the random perturbations. To unify physically distinct resources, we decompose this rate of change into \emph{intrinsic} and \emph{control-induced} components:
\begin{align}
	v(x,t) = \underbrace{f(x)}_{\text{Intrinsic dynamics}} + \underbrace{\gamma\,u(x,t)}_{\text{Control dynamics}},\ \forall t\in\mathcal{T},
	\label{eq:drift_decomp}
\end{align}
where $u(x,t)$ denotes a state-dependent control policy: $u(x,t)>0$ corresponds to a power load (e.g., charging/heating) and $u(x,t)<0$ corresponds to a power injection (e.g., discharging/cooling). The function $f(\cdot)$ captures intrinsic dynamics (e.g., thermal leakage of TCLs), and $\gamma>0$ maps electrical power to the normalized-state rate of change.

\begin{table}[h]
	\vspace{-5mm}
	\caption{Parameter Mapping for Generalized Storage-Like Resources}
	\label{tab:mapping}
	\vspace{-3mm}
	\centering
	\footnotesize
	\renewcommand{\arraystretch}{1.05}
	\begin{tabular}{l c c }
		\toprule
		\textbf{Parameters} & \textbf{EV or BESS} & \textbf{TCL}  \\
		\midrule
		State $x$ & SoC & $(\theta(t) - \theta^{\mathrm{min}})/(\theta^{\mathrm{max}} - \theta^{\mathrm{min}})$ \\
		Intrinsic $f(x)$ & $0$ & $-\alpha(x-x^{\mathrm{amb}})$  \\
		Coeff. $\gamma$ & $1/E^{\mathrm{cap}}$ & $\eta^{\mathrm{cop}}/\left(C^{\mathrm{th}}(\theta^{\mathrm{max}} - \theta^{\mathrm{min}})\right)$ \\
		\bottomrule
	\end{tabular}
	\vspace{-2mm}
\end{table}

Table~\ref{tab:mapping} specifies the unified model for two representative classes of storage-like DERs. The state variable $x \in [0, 1]$ represents the normalized energy-buffer level: for electrochemical storage (EVs/BESS), it corresponds directly to the SoC, whereas for TCLs it denotes the indoor temperature normalized by the user's comfort range $[\theta^{\mathrm{min}}, \theta^{\mathrm{max}}]$ through $x(t) = (\theta(t) - \theta^{\mathrm{min}})/(\theta^{\mathrm{max}} - \theta^{\mathrm{min}})$. The intrinsic dynamic function $f(x)$ characterizes the inherent system dynamics in the idle state ($u=0$): it is zero for ideal batteries (EVs/BESS), while for TCLs it captures thermal leakage as $f(x) = -\alpha (x - x^{\mathrm{amb}})$, with $x^{\mathrm{amb}}$ denoting the similarly normalized ambient temperature. The conversion coefficient $\gamma$ scales physical power to normalized state velocity, given by the reciprocal of energy capacity ($1/E^{\mathrm{cap}}$) for batteries and derived from the coefficient of performance and thermal capacitance ($\eta^{\mathrm{cop}}/C^{\mathrm{th}}$) for TCLs.

\subsubsection{Constraints of Individuals}
Due to device limits, the control policy $u(x,t)$ is bounded by the power lower bound $P^{\mathrm{min}}$ and upper bound $P^{\mathrm{max}}$:
\begin{align}
	u^{\mathrm{min}} \le u(x,t) \le u^{\mathrm{max}},\  \forall x\in\mathcal{X}, \ \forall t\in\mathcal{T}.
\end{align}
Substituting into \eqref{eq:drift_decomp} yields bounds on the drift velocity:
\begin{subequations}
\begin{align}
	&v^{\mathrm{min}}(x) \le v(x,t) \le v^{\mathrm{max}}(x),\  \forall x\in\mathcal{X}, \ \forall t\in\mathcal{T},
	\label{eq:v_bound_state} \\
	&v^{\mathrm{min}/\mathrm{max}}(x)\triangleq f(x)+\gamma u^{\mathrm{min}/\mathrm{max}},\  \forall x\in\mathcal{X}, \ \forall t\in\mathcal{T}.
	\label{eq:v_minmax_def}
\end{align}
\end{subequations}

\subsection{Macroscopic Density Evolution}
\subsubsection{State Evolution via Conservation Law}
\begin{figure}[H]
	\centering
	\vspace{-2mm}
	\includegraphics[width=1\columnwidth]{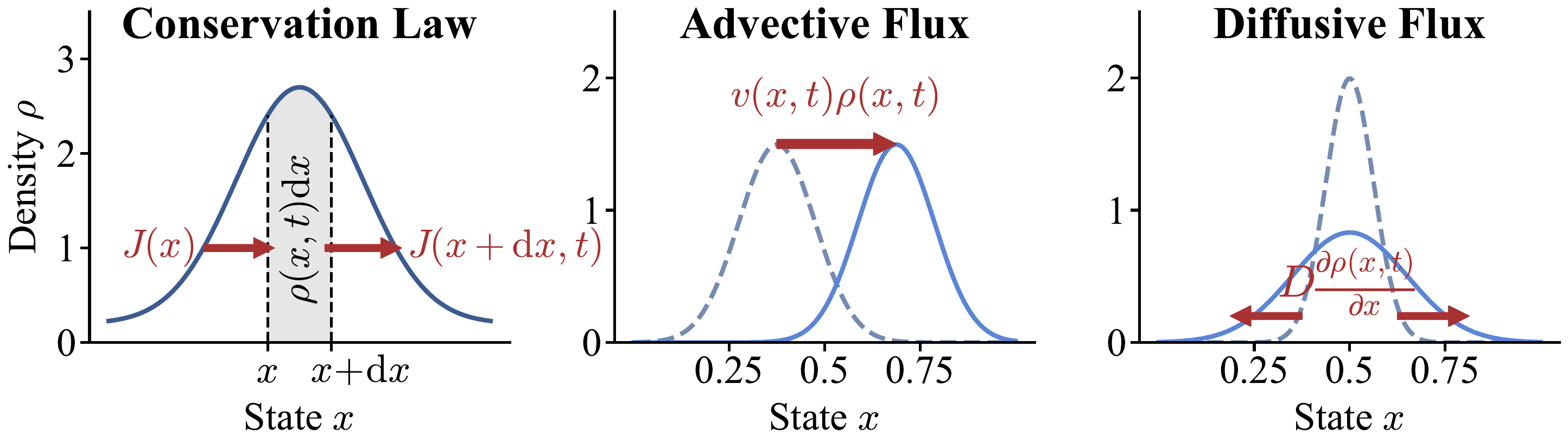}
	\vspace{-7mm}
	\caption{Schematic illustration of the physical mechanisms governing the macroscopic state evolution. \textbf{Left:} The principle of \emph{conservation of probability mass} over a differential control volume $[x, x+\mathrm{d}x]$. \textbf{Middle:} \emph{Advective flux} drives a coherent directional transport (shift) of the probability mass, representing the deterministic response to control signals. \textbf{Right:} \emph{Diffusive Flux} induces a spontaneous spreading of the distribution, representing the impact of resource-level uncertainties.}
	\label{fig:physics_mechanisms}
	\vspace{-2mm}
\end{figure}

As the population size $N \to \infty$, we approximate the aggregation as a continuum. Let $\rho(x,t)$ denote the state probability density. To derive its evolution, consider the probability-mass flow through a differential control volume $[x, x+\mathrm{d}x]$, as illustrated in Fig.~\ref{fig:physics_mechanisms}. By the law of \emph{conservation of probability mass}, the rate of change of the mass accumulated in this segment is determined by the difference between the inflow flux $J(x,t)$ and the outflow flux $J(x+\mathrm{d}x,t)$. Taking the limit as $\mathrm{d}x \to 0$ yields the continuity equation, i.e., the FPE:
\begin{align}
	\frac{\partial \rho(x,t)}{\partial t} + \frac{\partial J(x,t)}{\partial x} = 0, \ \forall x\in\mathcal{X}, \ \forall t\in\mathcal{T}. \label{eq:fpe_compact}
\end{align}
Here, $J(x,t)$ is the probability flux, which comprises two distinct physical mechanisms shown in Fig.~\ref{fig:physics_mechanisms}:
\begin{align}
	J(x,t) \triangleq \underbrace{v(x,t)\rho(x,t)}_{\text{Advective Flux}} - \underbrace{D \frac{\partial \rho(x,t)}{\partial x}}_{\text{Diffusive Flux}}. \label{eq:flux_def}
\end{align}
The \emph{Advective Flux} represents coherent transport driven by the deterministic control policy. Physically, it shifts the entire distribution directionally along the state space (e.g., charging moves the population toward higher energy states). The \emph{Diffusive Flux} captures spontaneous spreading caused by resource-level uncertainties. Following Fick's laws of diffusion \cite{1480898}, this term tends to smooth the density profile over time.
Finally, to confine the population within the physical limits $\mathcal{X} = [0, 1]$, we impose zero-flux boundary conditions:
\begin{align}
	J(x,t) \big|_{x\in\{0,1\}} = 0,\ \forall t\in\mathcal{T}, \label{eq:bc_compact}
\end{align}
which ensures that no probability mass leaks out of the feasible range of the state.

\subsubsection{Aggregate Power Calculation}
The aggregate power load $P^{\mathrm{agg}}(t)$ represents the collective contribution of the DER population. In the proposed decentralized architecture, each resource $i$ locally determines its power output $P_i(t)$ by evaluating the broadcast control policy $u(x,t)$ at its current state $x_i(t)$. By the law of large numbers, as $N \to \infty$, the sum of individual powers converges to an expectation over the macroscopic state density $\rho(x,t)$. Substituting the control relation $u(x,t)=(v(x,t)-f(x))/\gamma$ from \eqref{eq:drift_decomp}, the aggregate load can be written as:
\begin{align}
	P^{\mathrm{agg}}(t)
	&= \lim_{N\to\infty}\sum_{i=1}^{N} P_i(t) = N \int_{\mathcal{X}} u(x,t) \rho(x,t) \,\mathrm{d}x \notag \\
	&= E^{\mathrm{total}} \int_{\mathcal{X}} \big(v(x,t)-f(x)\big)\rho(x,t)\,\mathrm{d}x. \label{eq:p_agg_integral}
\end{align}
Here, $E^{\mathrm{total}} \triangleq N/\gamma$ denotes the \emph{total energy capacity} of the population under a homogeneous factor $\gamma$.\footnote{For populations with heterogeneous parameters (e.g., diverse battery capacities), this formulation can be generalized by defining $\rho(x, \gamma, t)$ as a joint density over the state and parameter spaces.} Physically, Eq. \eqref{eq:p_agg_integral} shows that the total grid load is determined by the difference between the macroscopic controlled evolution and the intrinsic dynamics, scaled by the total fleet capacity.

\subsubsection{Cyclic Constraint}
To ensure continuous dispatchability across consecutive dispatch horizons, we impose a hard cyclic constraint. This requires the terminal state distribution to strictly match the initial profile:
\begin{align}
	\rho(x,T)=\rho(x,0),
	\  \forall x\in\mathcal{X}.
	\label{eq:cyclic_hard}
\end{align}
Note the initial probability density $\rho(x,0)$ is acquired from the bottom up via the data mixing protocol introduced in Section~\ref{sec:overview}, so that the aggregator operates only on the collective distribution without accessing individual raw states.

\subsection{Formulation of the Aggregator Optimization}
We consider a price-taking aggregator that coordinates a flexible DER population under feeder-side exogenous base load and renewable generation conditions. The objective is to minimize the net electricity cost under the price $C(t)$, and the resulting optimization problem is formulated as:
\begin{subequations}
	\label{eq:primal_opt}
	\begin{align}
		\min_{\{v,\rho\},\,P^{\mathrm{g}}}~
		& \int_{0}^{T} C(t)\,P^{\mathrm{g}}(t)\,\mathrm{d}t
		\label{eq:primal_obj}
		\\
		\text{s.t.}~
		& P^{\mathrm{g,min}}\le P^{\mathrm{g}}(t)\le P^{\mathrm{g,max}},
		\quad \forall t,
		\label{eq:con_grid_limit}
		\\
		& P^{\mathrm{g}}(t)
		= P^{\mathrm{l}}(t)-P^{\mathrm{r}}(t)
		+ P^{\mathrm{agg}}(t),
		\ \forall t,
		\label{eq:con_coupling}
		\\
		& \text{Eqs. \eqref{eq:v_bound_state}-\eqref{eq:cyclic_hard}.}
	\end{align}
\end{subequations}
Constraint \eqref{eq:con_grid_limit} limits the net power exchange between the distribution grid and the aggregator, $P^{\mathrm{g}}(t)$, to remain within the allowable range $[P^{\mathrm{g,min}}, P^{\mathrm{g,max}}]$. Constraint \eqref{eq:con_coupling} enforces the power balance, where $P^{\mathrm{l}}$ and $P^{\mathrm{r}}$ represent the inelastic base load and renewable generation, respectively.
Directly solving the primal problem \eqref{eq:primal_opt} presents significant computational and theoretical challenges:
\begin{enumerate}[i)]
	\item \emph{Non-Convexity:} The advective flux term in \eqref{eq:flux_def} contains the bilinear product $v(x,t)\rho(x,t)$.
	\item \emph{Infinite Dimensionality:} Decision variables are functions defined over continuous spaces.
	\item \emph{Over-Conservatism of Exact Cyclicity:} Constraint \eqref{eq:cyclic_hard} enforces an exact distribution match. This rigidity precludes the opportunity to trade off small terminal deviations for significant operating cost reductions.
\end{enumerate}
These challenges motivate the tractable reformulation with a Wasserstein relaxation for near-cyclicity in Section~\ref{sec:solver}. 

\begin{remark}[Objective Generality]
	Although problem \eqref{eq:primal_opt} minimizes economic cost, the same structure can accommodate other convex aggregator-level tasks by replacing the objective, such as PCC power tracking, ramping penalties, feeder-interface regulation, or AGC signal tracking.
\end{remark}

\section{Sparse Linear Reformulation}
\label{sec:solver}

This section develops a convexification framework for reformulating the intractable PDE-constrained optimization problem \eqref{eq:primal_opt}. The key steps are: i) finite-volume discretization, which converts the continuous PDEs into algebraic difference equations; ii) a \emph{flux-lifting} substitution, which eliminates the bilinear non-convexity in the discretized formulation; and iii) a Wasserstein-based relaxation, which enables near-cyclic operation and improves economic flexibility. The resulting discretized model is a sparse LP that can be solved to global optimality by standard LP solvers.

\subsection{Finite-Volume Discretization}
\label{subsec:fvm}

We employ the finite-volume method (FVM) \cite{reddy2022finite} to discretize the state space. It integrates the PDE over control volumes, thereby preserving the discrete conservation of probability mass in the numerical model \cite{van2023experimental}.

\subsubsection{PDE Discretization}
We partition the normalized state space $\mathcal{X}=[0,1]$ into $K$ uniform cells (control volumes) of width $\Delta x=1/K$.
Let $\boldsymbol{\rho}_t = [\rho_{1,t}, \dots, \rho_{K,t}]^\top \in \mathbb{R}^K$ denote the vector of average densities at the cell centers.
Let $\boldsymbol{\Phi}_t = [\Phi_{1/2,t}, \dots, \Phi_{K+1/2,t}]^\top \in \mathbb{R}^{K+1}$ denote the vector of fluxes at the cell interfaces. The discretized FPE \eqref{eq:fpe_compact} enforces the mass-balance law: the density change in a cell equals the net inflow minus the net outflow:
\begin{align}
	\boldsymbol{\rho}_{t+1} = \boldsymbol{\rho}_{t} - \frac{\Delta t}{\Delta x} \mathbf{\Lambda} \boldsymbol{\Phi}_{t},\ \forall t\in\mathcal{T}, \label{eq:fvm_balance_derivation}
\end{align}
where $\mathbf{\Lambda} \in \mathbb{R}^{K \times (K+1)}$ is the incidence matrix mapping cell interfaces to cell centers. We further define the discrete divergence operator as $\mathbf{M} \triangleq \frac{\Delta t}{\Delta x}\mathbf{\Lambda}$ for notation brevity (see Appendix~\ref{app:fvm_ops} for the element-wise definition of $\mathbf{M}$).

The total interface flux $\boldsymbol{\Phi}_t = \boldsymbol{\Phi}^{\mathrm{adv}}_t + \boldsymbol{\Phi}^{\mathrm{dif}}_t$ consists of advective and diffusive components. Following \eqref{eq:flux_def}:
\begin{align}
	\begin{cases}
		\Phi^{\mathrm{adv}}_{k+1/2,t} &= v_{k+1/2,t} \, \rho_{k+1/2,t}, \\
		\Phi^{\mathrm{dif}}_{k+1/2,t} &= -D \frac{\rho_{k+1,t} - \rho_{k,t}}{\Delta x},
	\end{cases} \ \forall k\in\mathcal{K}. \label{eq:dif_flux_def}
\end{align}
To ensure numerical stability, we adopt an implicit-explicit scheme. The advective flux is treated explicitly (at time $t$), whereas the diffusive flux is treated implicitly (at time $t+1$) to stabilize the second-order derivative.
Substituting \eqref{eq:dif_flux_def} into \eqref{eq:fvm_balance_derivation} yields the discrete dynamics:
\begin{align}
	\boldsymbol{\rho}_{t+1} + \mathbf{M} \boldsymbol{\Phi}^{\mathrm{dif}}_{t+1} = \boldsymbol{\rho}_{t} - \mathbf{M} \boldsymbol{\Phi}^{\mathrm{adv}}_{t}. \label{eq:imex_intermediate}
\end{align}
Since $\boldsymbol{\Phi}^{\mathrm{dif}}_{t+1}$ is a linear function of $\boldsymbol{\rho}_{t+1}$, we can define a tridiagonal diffusion matrix $\mathbf{L} \in \mathbb{R}^{K \times K}$ such that $\mathbf{M}\boldsymbol{\Phi}^{\mathrm{dif}}_{t+1} = -\mathbf{L}\boldsymbol{\rho}_{t+1}$. This leads to the compact linear update rule:
\begin{align}
	(\mathbf{I} - \mathbf{L}) \boldsymbol{\rho}_{t+1} = \boldsymbol{\rho}_{t} - \mathbf{M} \boldsymbol{\Phi}^{\mathrm{adv}}_{t}. \label{eq:imex_final_form}
\end{align}
The zero-flux boundary conditions are enforced by fixing the boundary advective fluxes:
\begin{align}
	\Phi^{\mathrm{adv}}_{1/2,t}=0,\ \Phi^{\mathrm{adv}}_{K+1/2,t}=0, \quad \forall t\in\mathcal{T}. \label{eq:final_lp_advbc}
\end{align}

\subsubsection{Aggregate Power Discretization}
The aggregate load integral \eqref{eq:p_agg_integral} is discretized using the midpoint rule. Since $\rho$ is defined at cell centers and $\Phi^{\mathrm{adv}}$ at interfaces, we define matrix $\mathbf{A}^{\mathrm{avg}} \in \mathbb{R}^{K \times (K+1)}$ to map interface fluxes to cell centers. The load equation becomes purely linear in $\boldsymbol{\Phi}^{\mathrm{adv}}$ and $\boldsymbol{\rho}$:
\begin{align}
	P^{\mathrm{agg}}_{t} = E^{\mathrm{total}} \Delta x \mathbf{1}^{\top} \left( \mathbf{A}^{\mathrm{avg}} \boldsymbol{\Phi}^{\mathrm{adv}}_{t} - \mathbf{f} \circ \boldsymbol{\rho}_{t} \right), \label{eq:pagg_discrete}
\end{align}
where $\circ$ denotes the element-wise product and $\mathbf{f} \in \mathbb{R}^K$ is the vector of intrinsic dynamics at cell centers.

\subsection{Linearization via Flux Lifting}
\label{subsec:flux_lifting}

The primary computational challenge in the discretized model is the non-convex bilinear product $\Phi^{\mathrm{adv}} = v \cdot \rho$. To recover linearity, we employ a change-of-variable technique known as \emph{flux lifting}. The core idea is to treat the advective flux vector $\boldsymbol{\Phi}^{\mathrm{adv}}$ directly as the primary decision variable, eliminating variable $v$ from the optimization formulation.
Consequently, the original limits, $v^{\min}(x) \le v(x,t) \le v^{\max}(x)$, must be reformulated as constraints on the flux. By multiplying these inequalities by the non-negative density $\rho(x,t)$, we derive the following linear polyhedral constraints:
\begin{align}
	\mathbf{V}^{\mathrm{min}} \boldsymbol{\rho}_{t} \le \mathbf{A}^{\mathrm{avg}}\boldsymbol{\Phi}^{\mathrm{adv}}_{t} \le \mathbf{V}^{\mathrm{max}} \boldsymbol{\rho}_{t}, \quad \forall t \in \mathcal{T}, \label{eq:flux_polytope}
\end{align}
where $\mathbf{V}^{\mathrm{min}/\mathrm{max}}$ are diagonal matrices representing the limits of advective fluxes. Note that matrix $\mathbf{A}^{\mathrm{avg}}$ is applied to $\boldsymbol{\Phi}^{\mathrm{adv}}_{t}$ to map the fluxes from cell interfaces to cell centers, ensuring dimension consistency with the density $\boldsymbol{\rho}_{t}$.

\subsection{Wasserstein Relaxation for Near-Cyclicity}
To address the over-conservatism of pointwise cyclicity, we relax \eqref{eq:cyclic_hard} using the Wasserstein distance. For 1D distributions, this distance is equivalent to the $L_1$-norm of the difference between cumulative distribution functions (CDFs) \cite{deangelis20211wassersteindistanceareamarginal}.
Let $\mathbf{F}_t = \mathbf{S} \boldsymbol{\rho}_t$ be the discrete CDF of the state distribution, where $\mathbf{S} \in \mathbb{R}^{K \times K}$ is a lower-triangular summation matrix. The cyclic constraint \eqref{eq:cyclic_hard} is relaxed as:
\begin{align}
	W_1(\boldsymbol{\rho}_{T},\boldsymbol{\rho}_{0})
	= \| \mathbf{F}_T - \mathbf{F}_0 \|_1 \le \varepsilon^{\mathrm{cyc}}. \label{eq:w1_ball}
\end{align}
Parameter $\varepsilon^{\mathrm{cyc}} \ge 0$ is a tunable tolerance. It quantifies the permissible deviation from the initial state, effectively serving as a ``flexibility budget" that allows the aggregator to trade off exact state restoration for reduced operating costs.

To incorporate this into an LP, we introduce an auxiliary vector $\mathbf{z} \in \mathbb{R}^K$ to linearize the absolute value operator:
\begin{align}
	-\mathbf{z} \le \mathbf{S}(\boldsymbol{\rho}_T - \boldsymbol{\rho}_0) \le \mathbf{z}, \quad \Delta x \mathbf{1}^{\top}\mathbf{z} \le \varepsilon^{\mathrm{cyc}}. \label{eq:wasserstein_lp}
\end{align}

\subsection{Sparse Linear Program Formulation}
\label{subsec:final_lp}

By letting $\Xi=\{\boldsymbol{\rho}, \boldsymbol{\Phi}^{\mathrm{adv}}, \mathbf{P}^{\mathrm{g}}, \mathbf{z}\}$, problem \eqref{eq:primal_opt} is finally formulated as the following sparse LP: 
\begin{align}
	\min_{\Xi} \quad & \sum_{t\in\mathcal{T}} C_t P^{\mathrm{g}}_t\,\Delta t \label{eq:lp_obj} \\
	\text{s.t.} \quad 	
	& \text{Grid \& Balance: \eqref{eq:con_grid_limit}-\eqref{eq:con_coupling},} \notag \\
	& \text{Linear Dynamics: \eqref{eq:imex_final_form} \& \eqref{eq:final_lp_advbc},} \notag \\
	& \text{Aggregate Power: \eqref{eq:pagg_discrete},} \notag \\
	& \text{Flux Constraints: \eqref{eq:flux_polytope},} \notag \\
	& \text{Near-Cyclicity: \eqref{eq:wasserstein_lp},} \notag \\
	& \text{Probability Validity: } \boldsymbol{\rho}_{t}\ge \bm 0, \ \mathbf{1}^{\top}\boldsymbol{\rho}_{t}\,\Delta x = 1. \notag
\end{align}
The coefficient matrices ($\mathbf{I}-\mathbf{L}$, $\mathbf{M}$, $\mathbf{A}^{\mathrm{avg}}$) are highly sparse, enabling efficient factorization.

\begin{remark}[Population-Size Scalability]
	The number of variables and constraints in the LP \eqref{eq:lp_obj} are governed by the discretization fidelity $(|\mathcal{T}|,K)$ rather than the population size $N$. Hence, the solving time is insensitive to $N$.
\end{remark}

\section{Microscopic Execution}
\label{sec:dispatch}

Upon obtaining the optimal macroscopic solution $\boldsymbol{\rho}^\star$ and $(\boldsymbol{\Phi}^{\mathrm{adv}})^\star$ from the LP \eqref{eq:lp_obj}, the aggregator constructs a unified state-dependent control signal and broadcasts it to the population. Each individual resource then translates the signal into a specific power setpoint $P_i$ locally based on its state, without requiring the aggregator to track individual devices.

\subsection{State-Dependent Signal Reconstruction}
The aggregator first derives the optimal control signal $\mathbf{v}^\star(t) \in \mathbb{R}^K$ from the macroscopic solution $(\boldsymbol{\rho}^\star, (\boldsymbol{\Phi}^{\mathrm{adv}})^\star)$. Based on the advective flux definition \eqref{eq:dif_flux_def}, the control signal at cell centers is computed as:
\begin{align}
	v_{k}^\star(t) = \frac{\big(\mathbf{A}^{\mathrm{avg}}(\boldsymbol{\Phi}^{\mathrm{adv}}_t)^\star\big)_k}{\boldsymbol{\rho}^\star_{k,t} + \sigma}, \ \forall k\in\mathcal{K}, \ \forall t\in\mathcal{T}.
	\label{eq:recover_v}
\end{align}
where $\sigma > 0$ (e.g., $10^{-8}$) is a regularization parameter introduced to ensure numerical stability in unoccupied cells (where $\boldsymbol{\rho}^\star_{k,t} \approx 0$). Meanwhile, matrix $\mathbf{A}^{\mathrm{avg}}$ is applied here to project the interface-defined fluxes onto cell centers, ensuring spatial alignment with the density $\boldsymbol{\rho}^\star$.

\subsection{Decentralized Execution by Individual Resources}
Each resource $i$ first locates its position within the discretized mesh by determining the cell index $k_i(t) = \lceil x_i(t)/\Delta x\rceil$. It then retrieves the target control value $v_{k_i}^\star(t)$ from the broadcast signal and identifies the corresponding intrinsic dynamics $f_{k_i(t)}$ (defined as the value of $f(x)$ at the center of cell $k_i(t)$). By inverting the control-affine dynamics \eqref{eq:drift_decomp}, the resource calculates its local power setpoint $u_{i}(t)$ as:
\begin{align}
	u_{i}(t) = \frac{1}{\gamma_i} \left( v_{k_i(t)}^\star(t) - f_{k_i(t)} \right), \quad \forall i \in \mathcal{N}, \forall t \in \mathcal{T}.
	\label{eq:local_response}
\end{align}
\begin{remark}[Recovered-Signal Feasibility on Occupied Cells]
	For cells with non-negligible density, if $\rho^\star_{k,t}\ge \rho^{\min}>0$ and $\sigma \to 0$, then the flux bounds in \eqref{eq:flux_polytope} imply
	\begin{align}
		v^{\min}_{k} \le \frac{\big(\mathbf{A}^{\mathrm{avg}}(\boldsymbol{\Phi}^{\mathrm{adv}}_t)^\star\big)_k}{\rho^\star_{k,t}} \le v^{\max}_{k},
	\end{align}
	so the reconstructed local setpoint obtained from \eqref{eq:local_response} lies within $[u^{\min},u^{\max}]$. For cells with vanishing density, the regularization term $\sigma$ is introduced only for numerical robustness; in implementation, the reconstructed signal is clipped to $[v^{\min}_{k},v^{\max}_{k}]$ before being mapped to local setpoints. Therefore, the residual violations observed in Monte-Carlo simulations are attributable to discretization error, stochastic diffusion, and reconstruction regularization/clipping, rather than infeasibility of the LP itself.
\end{remark}


\section{Case Study} \label{sec:case}

\subsection{Simulation Setup}
\label{subsec:case_setup}

\subsubsection{Common Parameter Settings}
\begin{figure}[H]
	\vspace{-4mm}
	\subfigcapskip=-2pt
	\centering
	\subfigure[Electricity Prices]{\includegraphics[width=0.49\columnwidth]{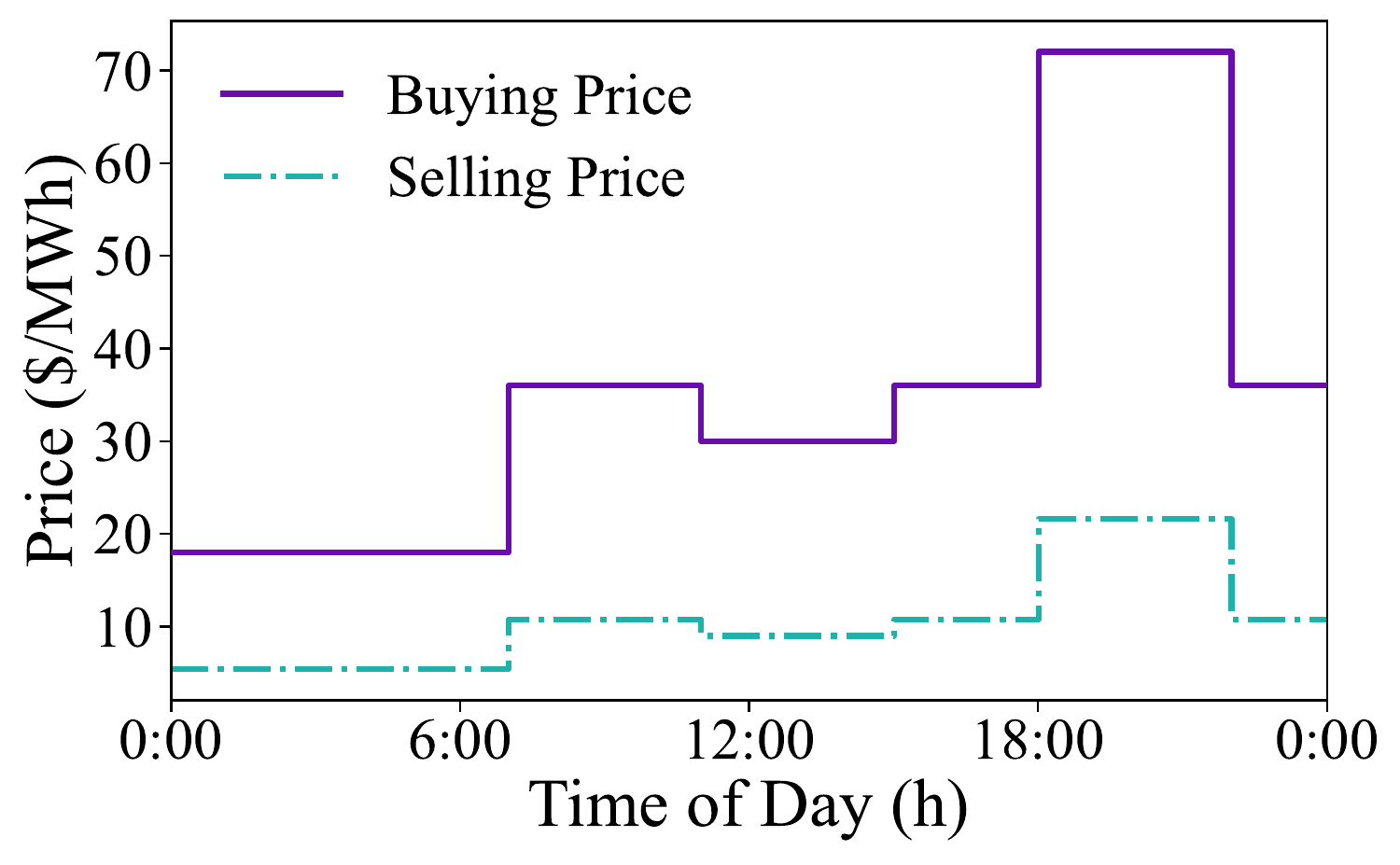}}
	\subfigure[Load and Generation]{\includegraphics[width=0.49\columnwidth]{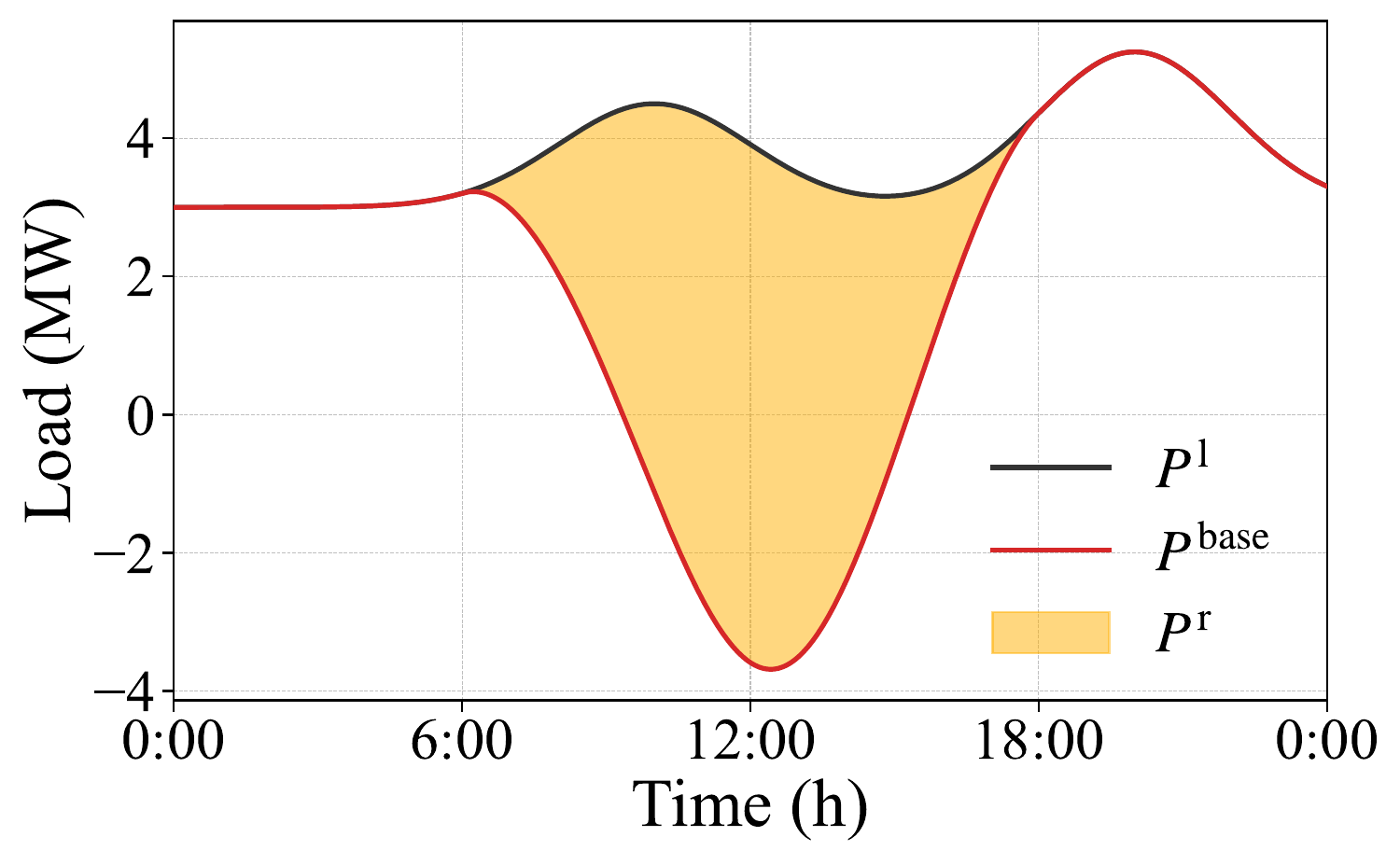}}
	\vspace{-4mm}
	\caption{Prices for purchasing and selling electricity, together with the base-load (excluding the DER population), renewable-generation, and base net-load (base load minus renewable generation) profiles for a population of 1,000 EVs.}
	\label{fig_Parameters}
	\vspace{-4mm}
\end{figure}

This section evaluates the proposed Eulerian coordination framework in two representative \emph{single-population} settings: an EV population (Case~I) and a TCL population (Case~II), representing nearly lossless and leaky storage-like flexibility, respectively. This pairing is intentional: together, the two cases cover both zero-drift and nonzero-drift resource physics and therefore test whether the framework remains effective across distinct forms of intertemporal flexibility. The electricity prices, base load (excluding the DER population), and renewable generation profiles are shown in Fig.~\ref{fig_Parameters}; the detailed population parameters are given in the corresponding subsections.

\subsubsection{Benchmark}
We benchmark the proposed method against a \emph{centralized} device-level Lagrangian model that explicitly optimizes individual trajectories under the same exogenous price, base-load, renewable-generation, and PCC exchange settings. This benchmark enforces individual hard cyclicity, i.e., each device must satisfy $x_i(T)=x_i(0)$, and therefore serves as a strong reference for economic performance and schedule structure. Its per-device cyclicity is stricter than the distribution-level cyclicity in the proposed formulation, so the reported cost gaps should be interpreted as empirical proximity to a stronger baseline rather than exact optimality under an identical cyclicity definition.

\subsubsection{Evaluation Procedure}
All reported metrics are obtained via Monte-Carlo simulation. After solving the macroscopic LP and broadcasting the dispatch signal, each device evolves according to the microscopic SDE in \eqref{eq:micro_sde}, including the stochastic diffusion term. We then compute the realized population-level operating costs, solving times, state-bound violations, and cyclic deviations. This procedure is important because it evaluates not only the macroscopic optimizer itself, but also the practical fidelity of the full two-layer pipeline after stochastic decentralized execution.

\subsection{Case I: Coordination of EV Population}
\label{subsec:case_ev}

\subsubsection{Default Parameter Settings}
\label{subsubsec:ev_params}

We consider an EV population participating in the aggregator self-scheduling problem. Unless otherwise stated, the default parameters are $N=1000$, $T=24$~h, $\Delta t=1$~min, and $K=200$ for the normalized SoC space $\mathcal{X}=[0,1]$. The per-EV energy capacity is $E_i^{\mathrm{cap}}=\mathrm{60}$~kWh, the charging/discharging limits are $P^{\min}=-7$~kW and $P^{\max}=7$~kW, and the diffusion coefficient in \eqref{eq:micro_sde} is $D=10^{-3}$. The initial SoC distribution is induced by $x_i(0)\sim \mathbb{N}(0.4,0.1^2)$, the near-cyclicity tolerance is set to $\varepsilon^{\mathrm{cyc}}=0$, and the PCC exchange limits are $P^{\mathrm{g,min}}=\mathrm{-5.6}$~MW and $P^{\mathrm{g,max}}=\mathrm{5.6}$~MW.

\subsubsection{Evaluation under Varying Population Scales}

We first evaluate the proposed method under varying EV population scales and benchmark it against the device-level Lagrangian model to assess scalability and empirical performance. The base load, renewable generation, and grid-exchange limits are all scaled linearly with the EV population size to maintain consistent per-capita operating conditions across different $N$.

\begin{figure}
	\vspace{-4mm}
	\subfigcapskip=-2pt
	\centering
	\subfigure[Total Costs]{\includegraphics[width=0.48\columnwidth]{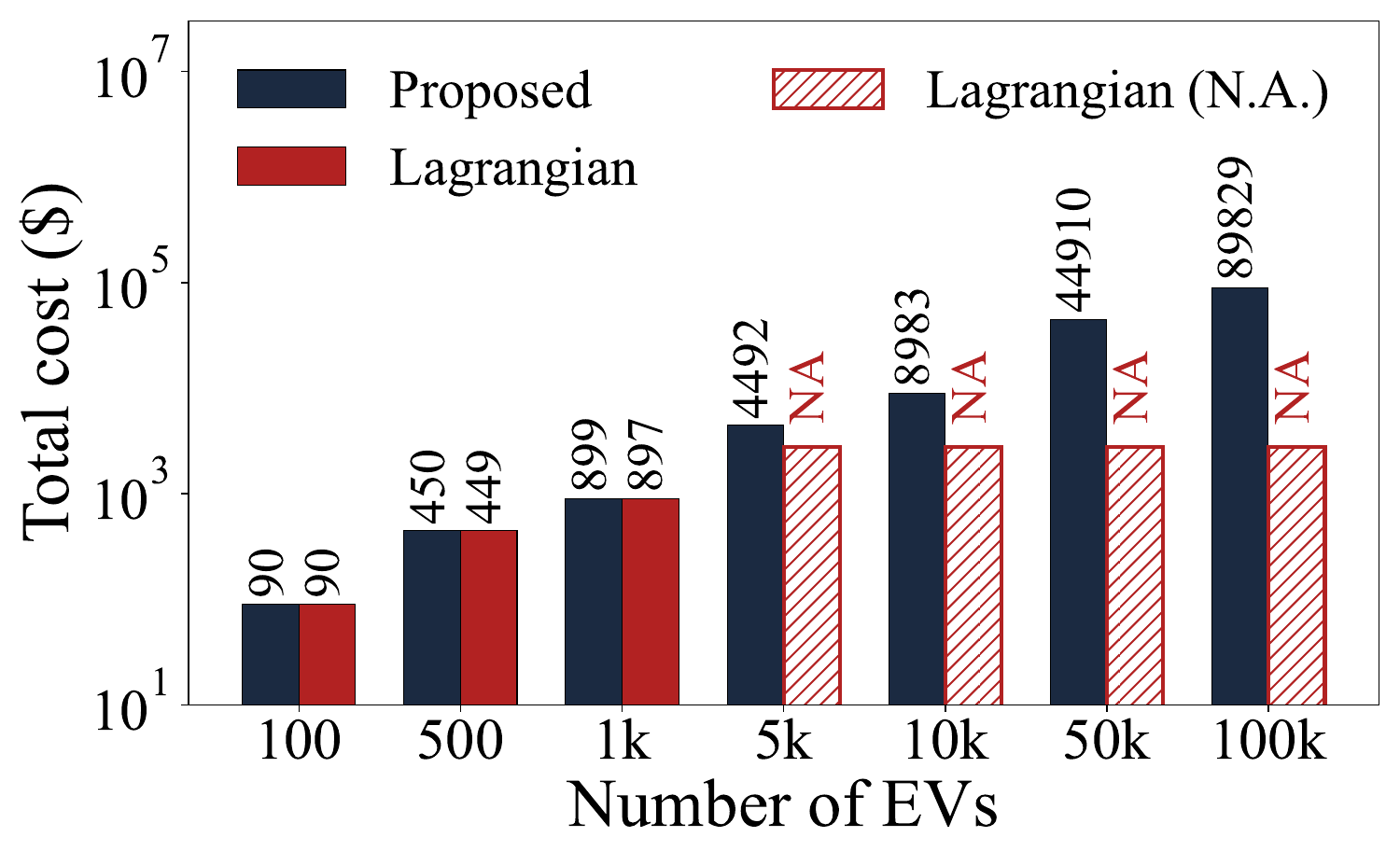}}
	\subfigure[Solving Times]{\includegraphics[width=0.5\columnwidth]{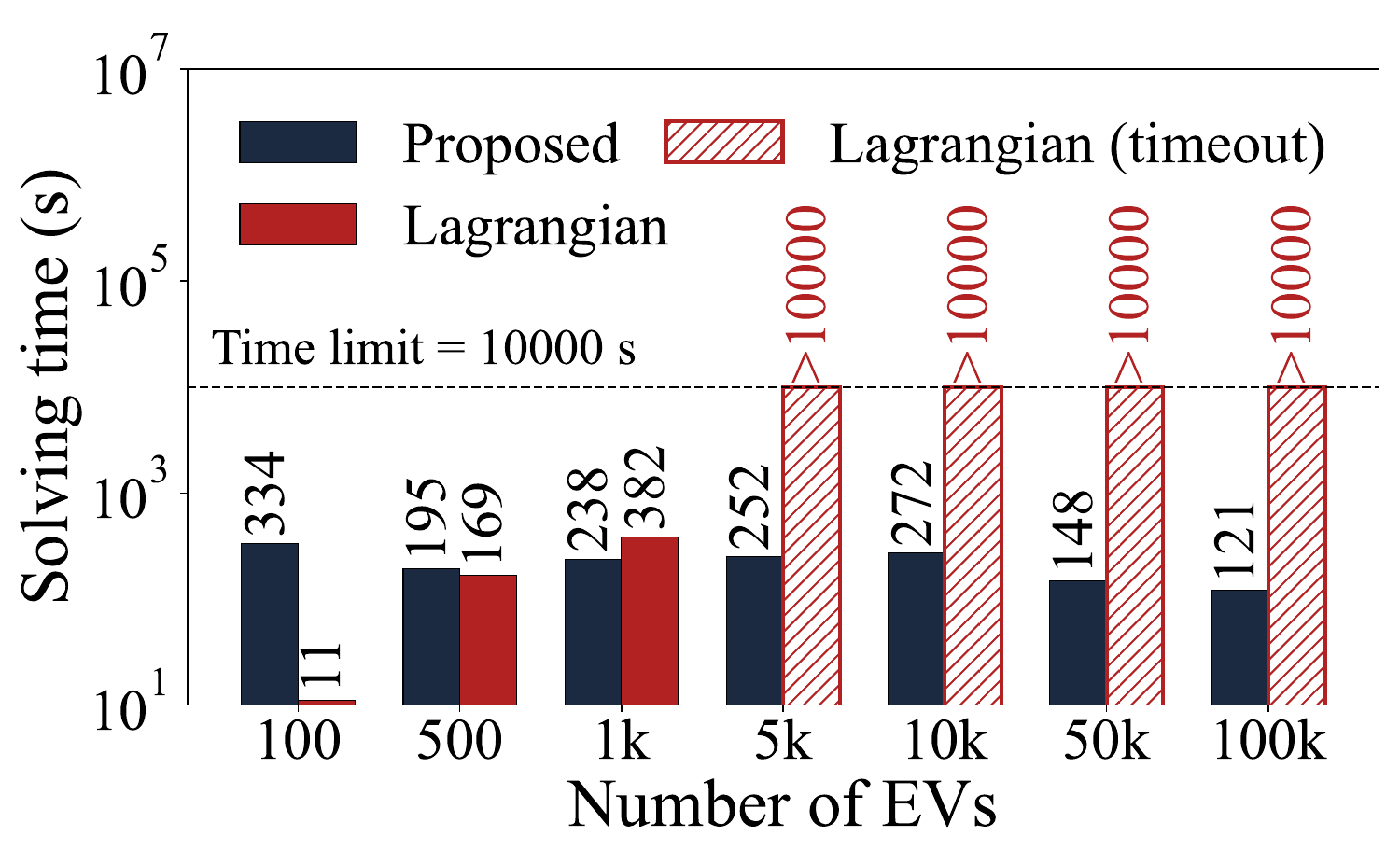}}
	\vspace{-4mm}
	\caption{Comparison of total costs and solving times between the proposed Eulerian method and the Lagrangian benchmark across varying EV population scales. Note that the Lagrangian benchmark results for populations of 5,000 EVs and above are omitted, as the Gurobi solver failed to find the optimal solutions within the 10,000 s time limit.}
	\label{fig_cost_EV_different_N}
	\vspace{-4mm}
\end{figure}

Fig.~\ref{fig_cost_EV_different_N} compares the total operating costs and solving times of the proposed Eulerian method and the Lagrangian benchmark under varying EV population scales. When the benchmark remains tractable (e.g., up to $N=1{,}000$), the proposed method achieves closely matching costs, with a relative gap consistently below $0.2\%$, despite the stricter per-device cyclicity imposed by the benchmark. In contrast, the Lagrangian benchmark rapidly becomes computationally prohibitive and fails to return optimal solutions within the $10{,}000$~s time limit for $N\ge 5{,}000$, while the proposed method remains efficient with solving times on the order of $10^2$~s. This trend is consistent with the fact that the macroscopic LP dimension is governed by the discretization rather than by the number of EVs.
This result is central to the proposed framework: it shows that moving from device-level trajectories to macroscopic density transport does not materially sacrifice economic performance in the tested regime, while it fundamentally changes how the computational burden scales with population size.

\begin{figure}
	\vspace{-4mm}
	\subfigcapskip=-2pt
	\centering
	\subfigure[Proposed Method]{\includegraphics[width=0.48\columnwidth]{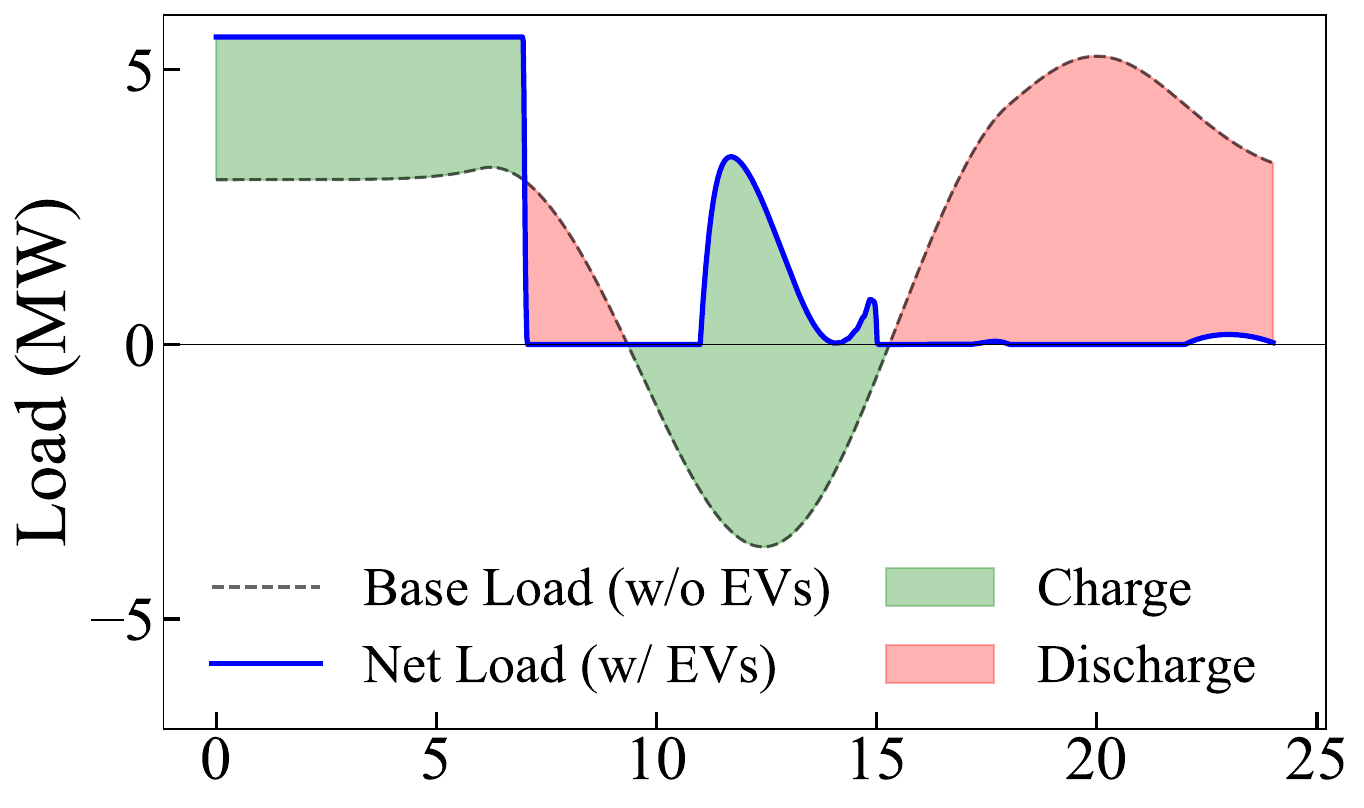}}
	\subfigure[Lagrangian Benchmark]{\includegraphics[width=0.5\columnwidth]{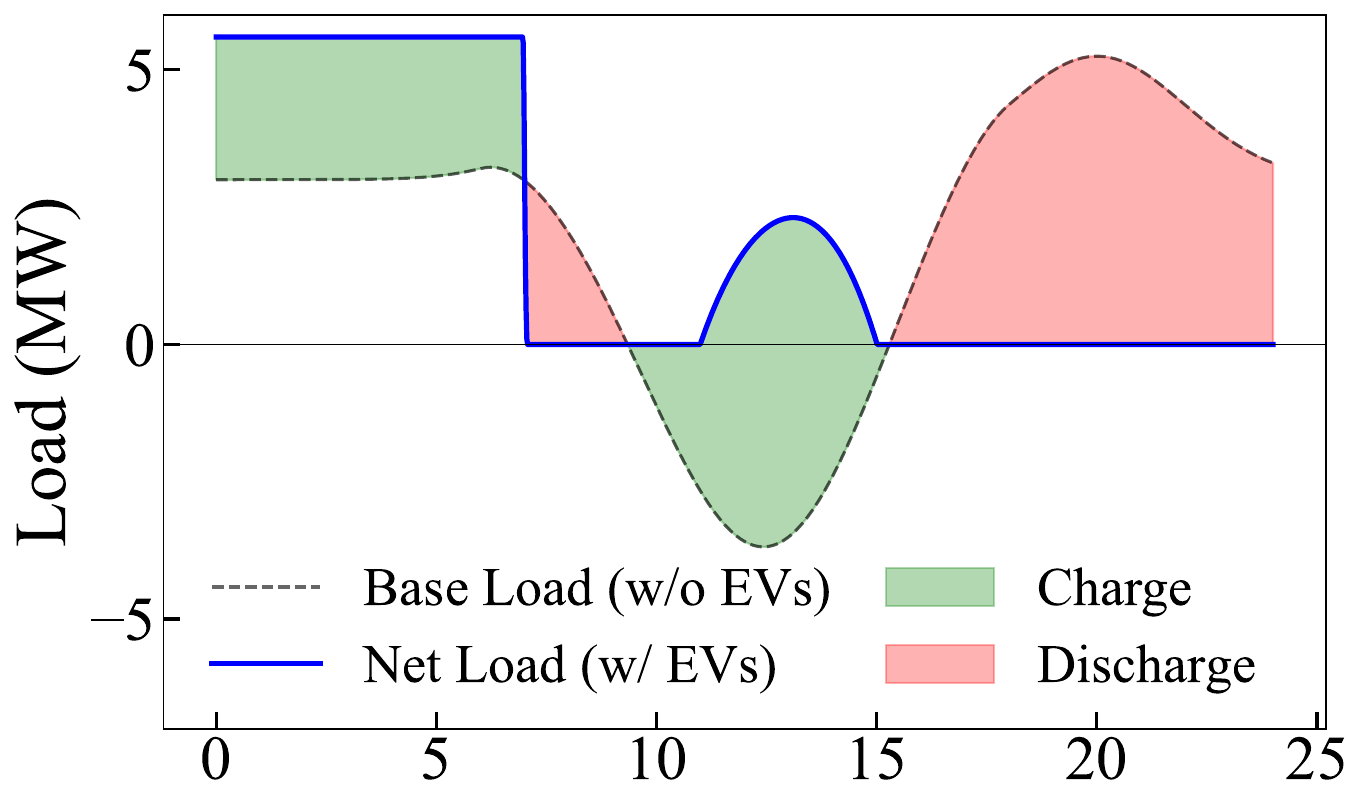}}
	\vspace{-4mm}
	\caption{Optimized load profiles of the proposed Eulerian method and the Lagrangian benchmark for a population of 1,000 EVs. The ``base load'' denotes the base system demand without considering EV loads, while the ``net load'' incorporates the optimized EV charging/discharging power.}
	\label{fig_profile_EV_different_N}
	\vspace{-4mm}
\end{figure}

Fig.~\ref{fig_profile_EV_different_N} illustrates the optimized load profiles for a representative case with $N=1{,}000$ EVs. The proposed method produces a net-load trajectory whose charging and discharging intervals closely match the tested device-level reference, consistent with the small cost gap in Fig.~\ref{fig_cost_EV_different_N}. This agreement indicates that the macroscopic formulation reproduces the dominant load-shaping behavior of the tested device-level benchmark in this representative large-population setting.

\begin{figure}[H]
	\vspace{-4mm}
	\subfigcapskip=-2pt
	\centering
	\subfigure[SoC Bound Constraint]{\includegraphics[width=0.49\columnwidth]{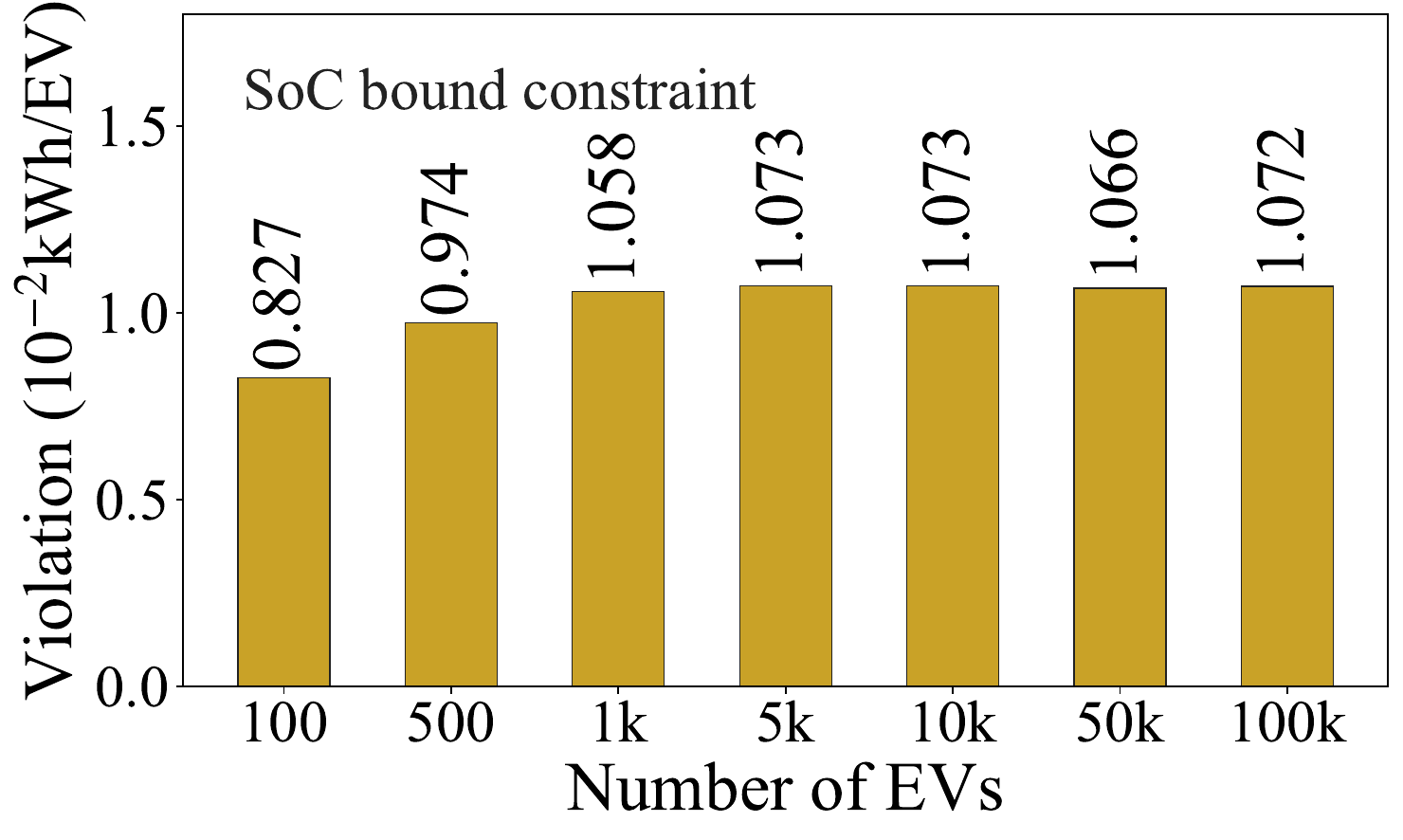}}
	\subfigure[Cyclic Constraint]{\includegraphics[width=0.49\columnwidth]{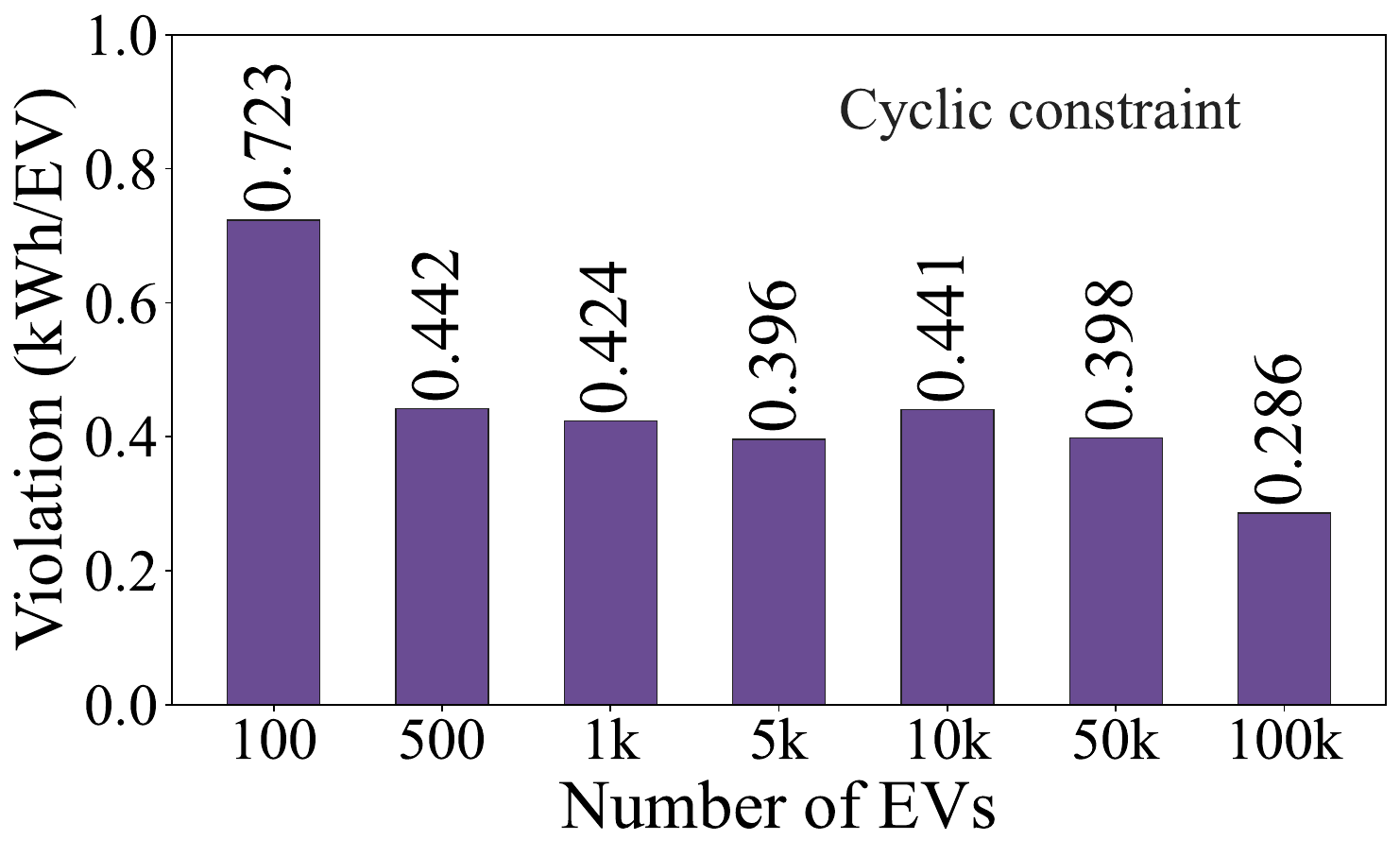}}
	\vspace{-4mm}
	\caption{Average violations of SoC bound and cyclic constraints per EV incurred by the proposed Eulerian method across varying population scales.}
	\label{fig_vio_EV_different_N}
	\vspace{-4mm}
\end{figure}

Fig.~\ref{fig_vio_EV_different_N} reports feasibility metrics of decentralized execution under Monte-Carlo simulations. The per-EV average violation is defined as the population average of each EV's \emph{cumulative} constraint violation over the 24-h horizon. The SoC-bound violation remains on the order of $10^{-2}$~kWh per EV, while the cyclic deviation also remains small and decreases markedly as the population grows (e.g., from $0.723$ to $0.286$~kWh/EV when $N$ increases from $100$ to $100{,}000$), consistent with improved fidelity of the Eulerian approximation.

\begin{figure}[H]
	\vspace{-4mm}
	\centering	{\includegraphics[width=1\columnwidth]{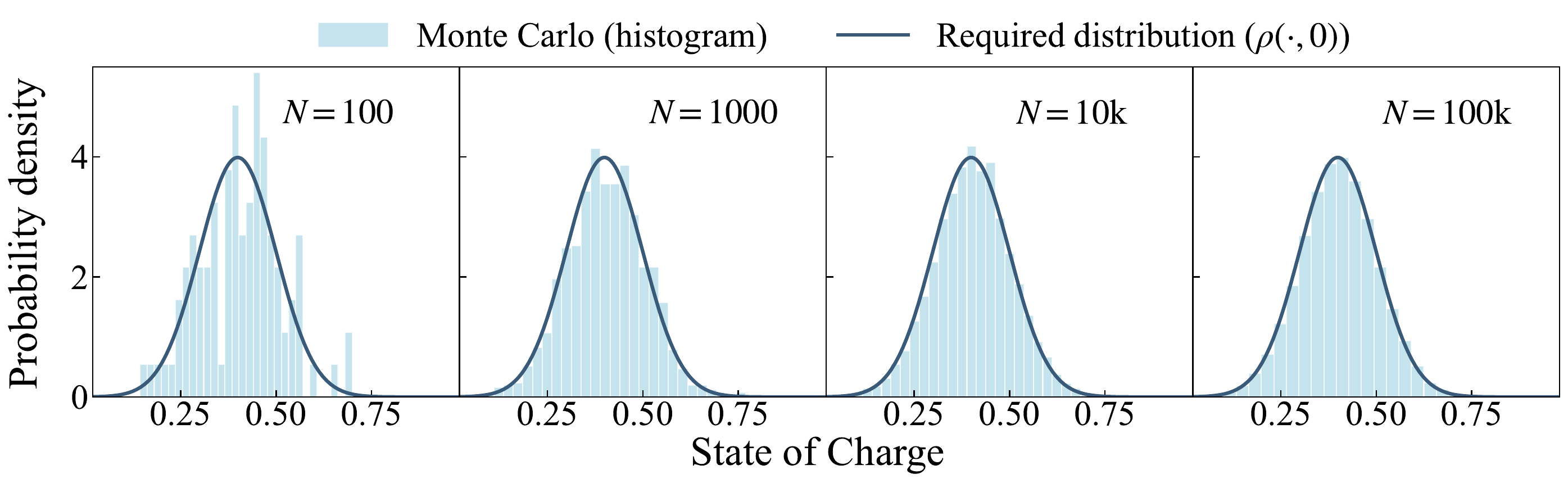}}
	\vspace{-8mm}
	\caption{Comparison between the target SoC distribution defined by the cyclic constraint (blue curve) and the realized distribution obtained via Monte-Carlo simulations (density histogram) across varying EV population scales.}
	\vspace{-4mm}
	\label{fig_cyclic_different_N}
\end{figure}

Fig.~\ref{fig_cyclic_different_N} provides a direct distributional validation of macroscopic-microscopic consistency. As $N$ increases, the realized terminal distributions increasingly align with the target profile, explaining the decreasing cyclic deviation in Fig.~\ref{fig_vio_EV_different_N}(b).

\subsubsection{Impact of Discretization Resolution}

The performance of the proposed framework depends on the accuracy of the discretized macroscopic model, which is governed by the temporal step size $\Delta t$ and the spatial resolution $K$ used in the finite-volume discretization. To examine the trade-off between accuracy and efficiency, we evaluate the proposed method under varying temporal and spatial resolutions.\footnote{In this sensitivity analysis, the Gurobi parameter \texttt{NumericFocus} is set to 3 (enabling quad-precision arithmetic) to mitigate potential numerical instability induced by varying discretization resolutions. While this ensures robust convergence, it inevitably leads to increased solving times and negligible deviations in the converged optima compared to the default precision setting.}

\begin{figure}[H]
	\vspace{-4mm}
	\subfigcapskip=-2pt
	\centering
	\subfigure[Cost Gaps]{\includegraphics[width=0.49\columnwidth]{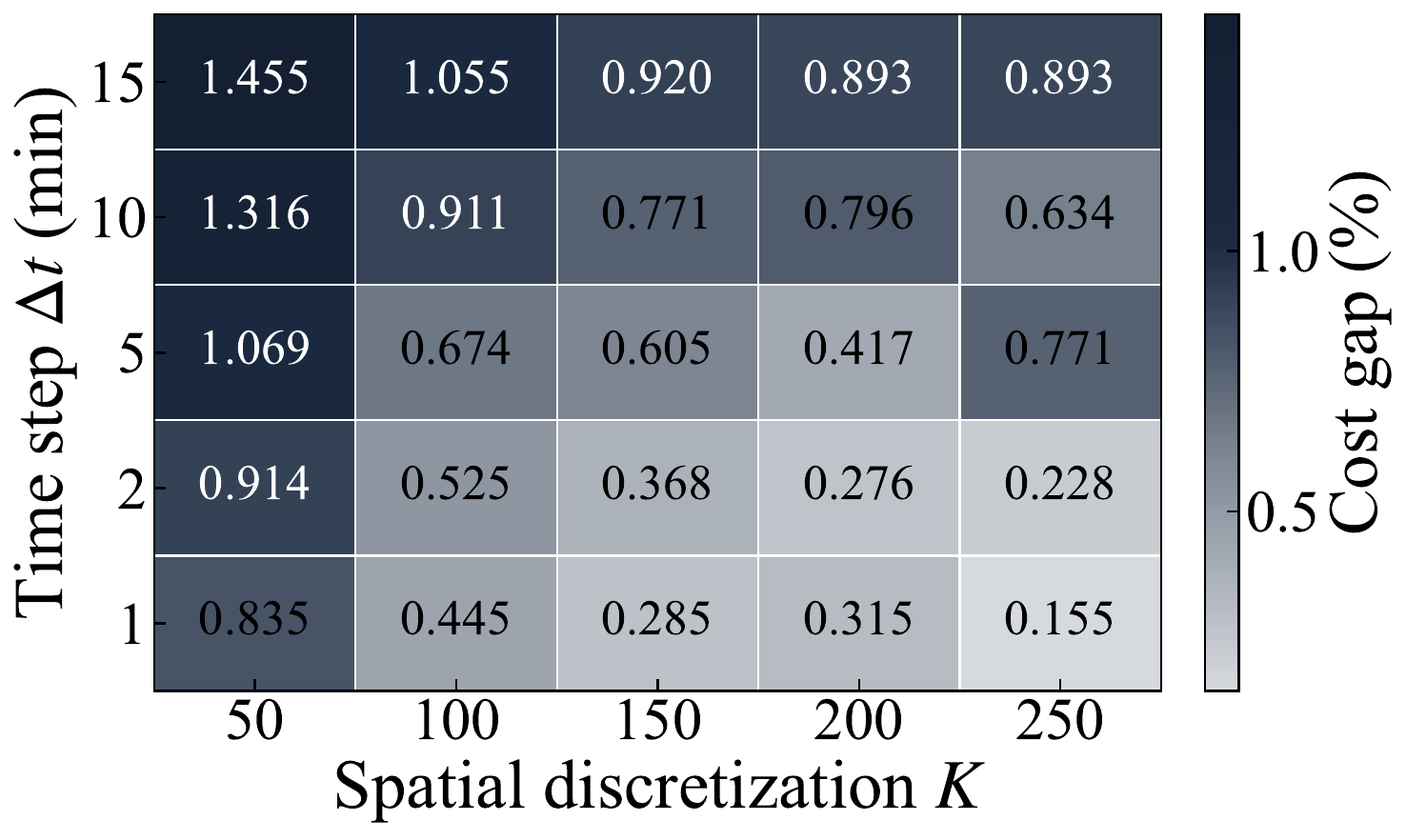}}
	\subfigure[Solving Times]{\includegraphics[width=0.49\columnwidth]{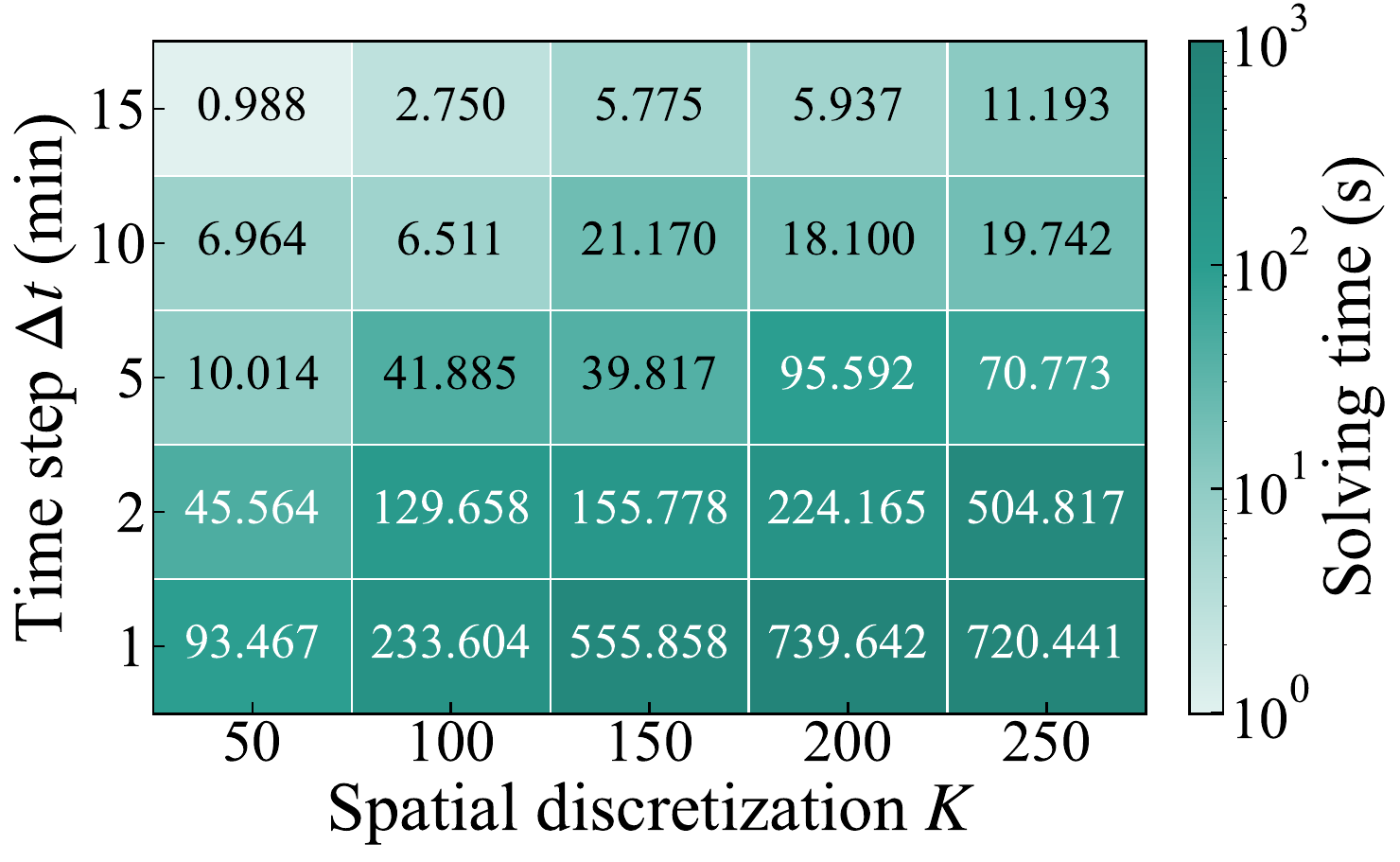}}
	\vspace{-4mm}
	\caption{Impact of temporal and spatial discretization resolutions on the cost gap (relative to the Lagrangian benchmark) and the solving time of the proposed method.}
	\label{fig_cost_EV_different_dtK}
	\vspace{-4mm}
\end{figure}

Fig.~\ref{fig_cost_EV_different_dtK} quantifies this trade-off by reporting the cost gap (relative to the Lagrangian benchmark) and the solving time across different $(\Delta t,K)$. As the discretization is refined, the cost gap generally decreases, e.g., from about $1.46\%$ at $(\Delta t=15~\mathrm{min},\,K=50)$ to about $0.16\%$ at $(\Delta t=1~\mathrm{min},\,K=250)$, at the expense of increased computation. The results suggest diminishing returns in cost-gap improvement at very fine resolutions while the computational burden continues to increase.
This behavior is consistent with the role of discretization in the model: finer temporal and spatial meshes provide a more faithful approximation of the underlying continuous density transport, but they also increase the size of the linear program and the numerical burden of the solver.

\begin{figure}[H]
	\vspace{-4mm}
	\subfigcapskip=-2pt
	\centering
	\subfigure[SoC Bound Constraint]{\includegraphics[width=0.49\columnwidth]{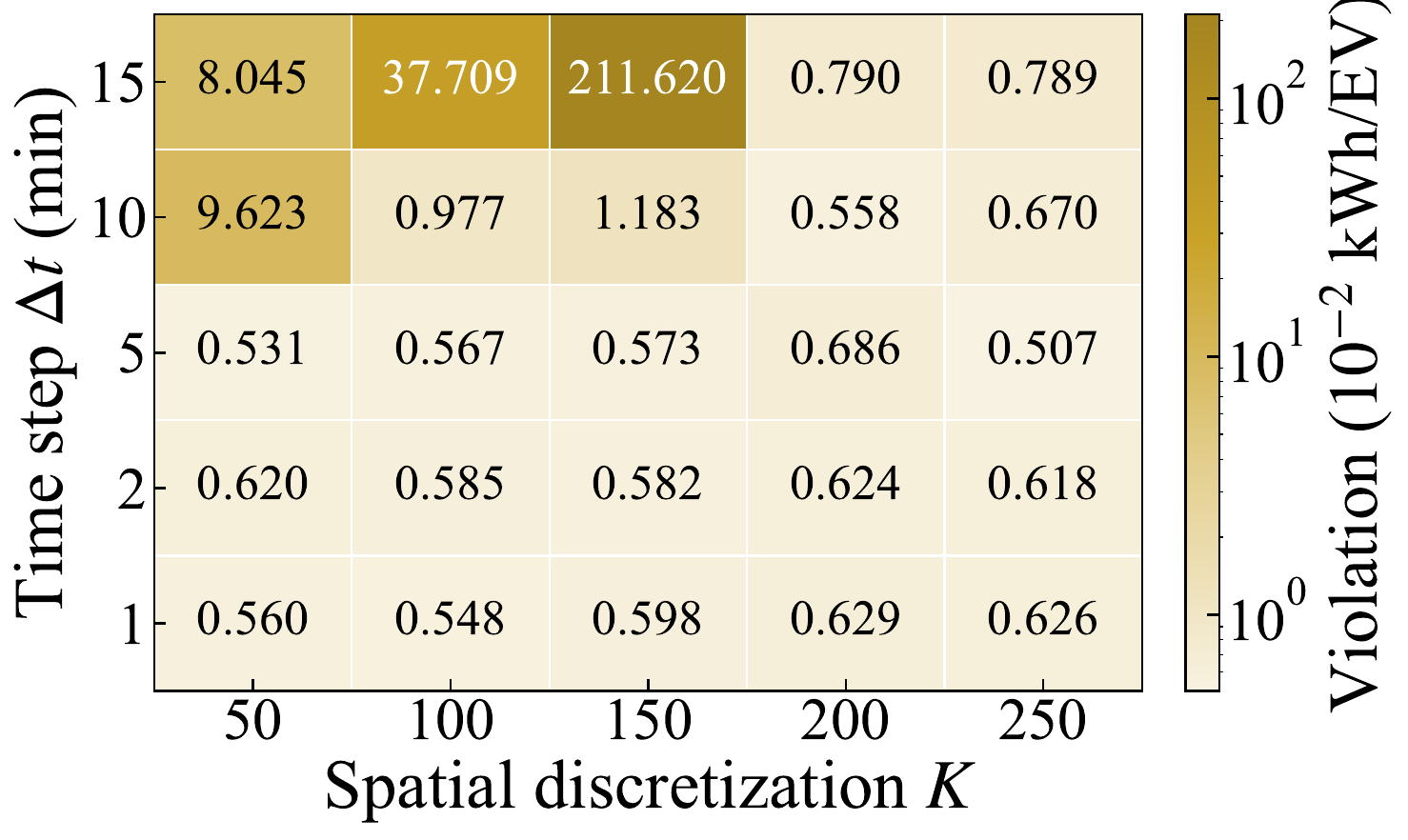}}
	\subfigure[Cyclic Constraint]{\includegraphics[width=0.49\columnwidth]{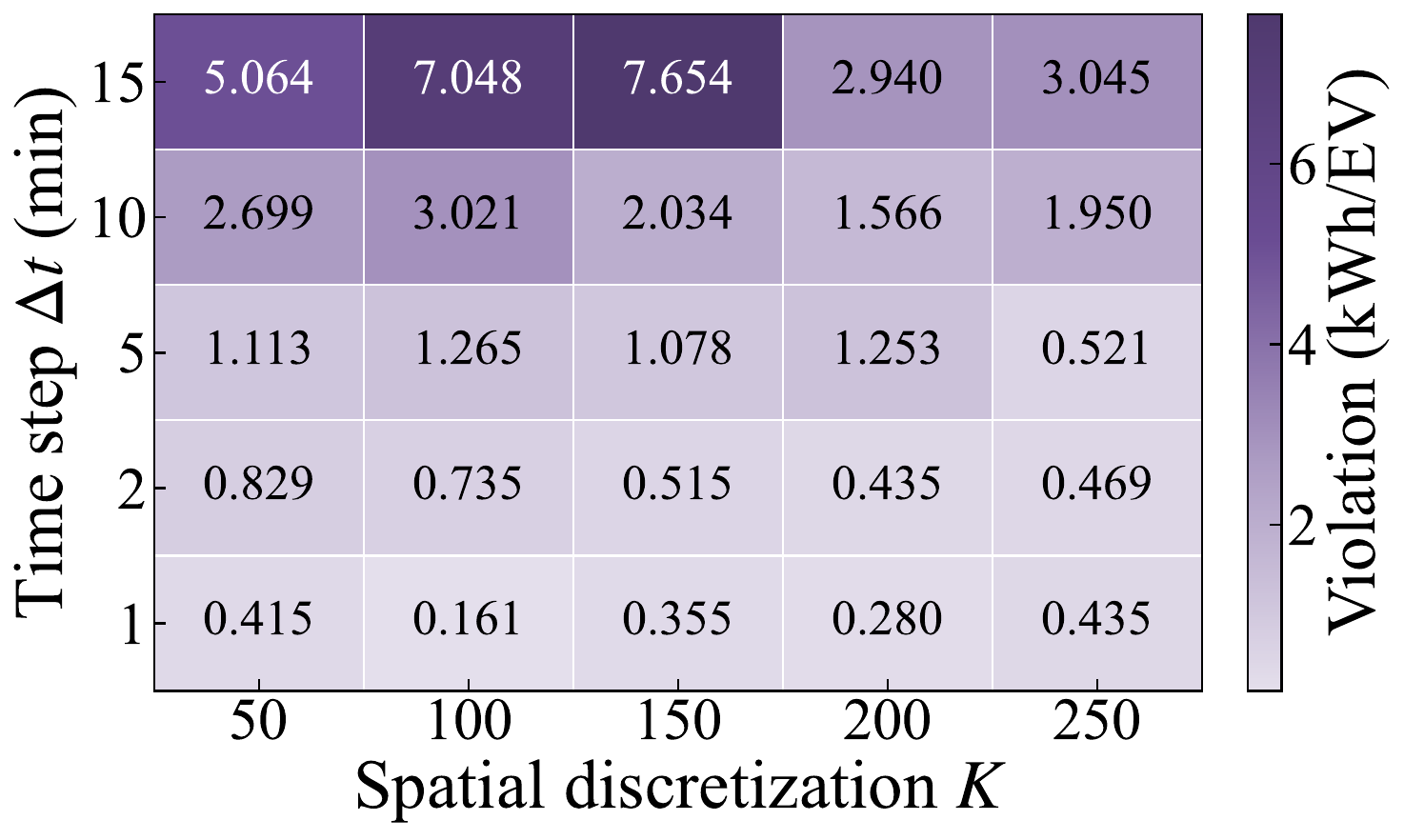}}
	\vspace{-4mm}
	\caption{Impact of temporal and spatial discretization resolutions on SoC bound and cyclic constraint violations under the proposed method.}
	\label{fig_vio_EV_different_dtK}
	\vspace{-4mm}
\end{figure}

Fig.~\ref{fig_vio_EV_different_dtK} reports feasibility metrics under Monte-Carlo decentralized execution. Across most resolutions, the SoC-bound violation remains on the order of $10^{-2}$~kWh/EV, whereas the cyclic deviation is more sensitive and decreases substantially as the resolution is refined. This is consistent with the fact that terminal-distribution matching amplifies accumulated approximation errors in time stepping and flux discretization.

\begin{figure}
	\vspace{-4mm}
	\centering	{\includegraphics[width=1.0\columnwidth]{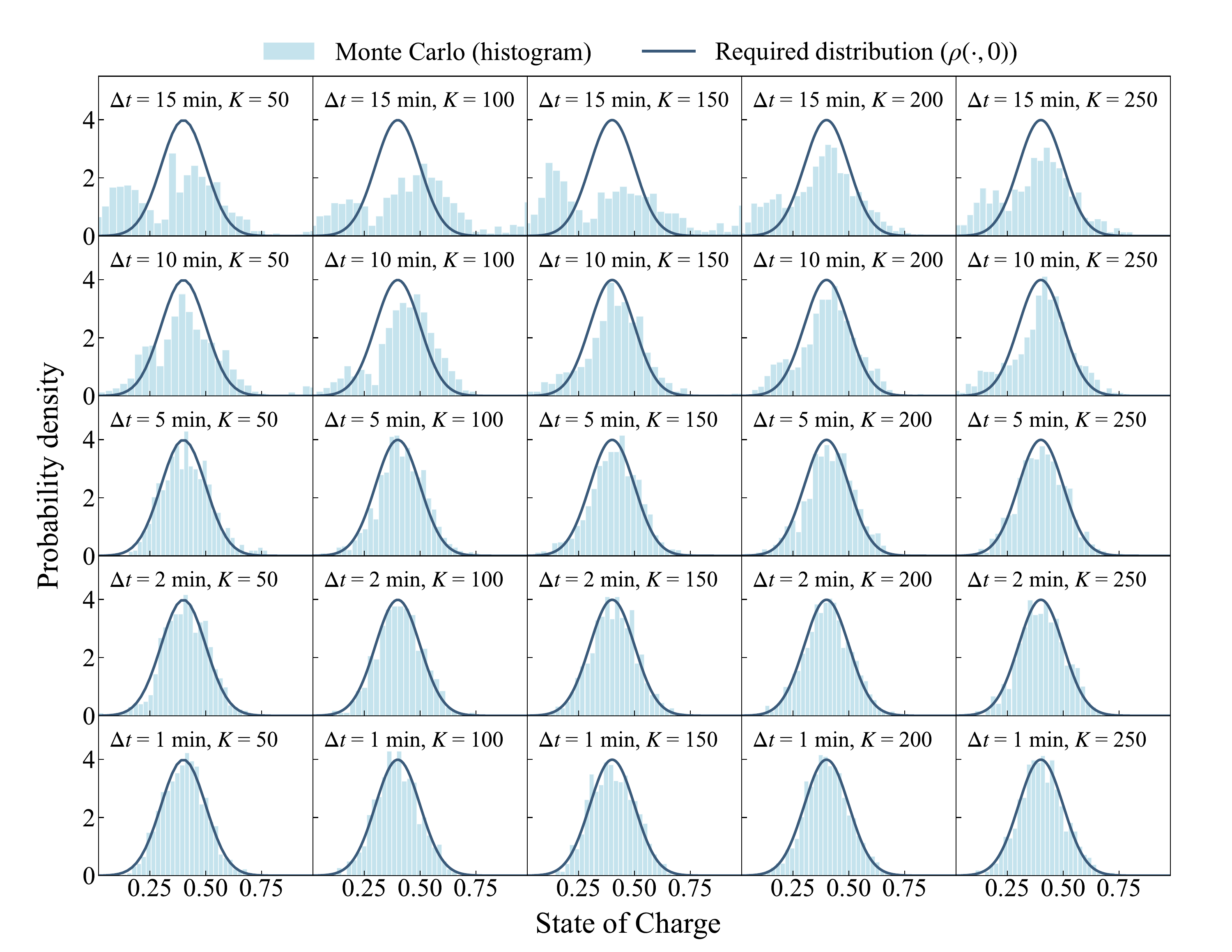}}
	\vspace{-8mm}
	\caption{Comparison between the target SoC distribution defined by the cyclic constraint (blue curve) and the realized distribution obtained via Monte-Carlo simulations (density histogram) across varying temporal and spatial discretization resolutions for the EV population.}
	\vspace{-4mm}
	\label{fig_cyclic_different_dtK}
\end{figure}

Fig.~\ref{fig_cyclic_different_dtK} visualizes the cyclicity realization under different resolutions. Visible mismatches occur under coarse resolutions, but the realized distributions increasingly align with the target profile as $\Delta t$ decreases and $K$ increases. Based on these results, the default setting $(\Delta t=1~\mathrm{min},\,K=200)$ offers a favorable trade-off between accuracy and efficiency.

\subsubsection{Impact of the Wasserstein Relaxation Radius}

While the hard cyclic requirement ($\varepsilon^{\mathrm{cyc}}=0$) restores the population energy buffer at the end of each horizon, it can restrict economic flexibility. To quantify this economics-cyclicity trade-off, we vary the Wasserstein relaxation radius $\varepsilon^{\mathrm{cyc}}$ in \eqref{eq:w1_ball} and evaluate both the operating cost and the realized deviation from cyclicity under Monte-Carlo decentralized execution. Here, the ``cyclic deviation'' is reported in kWh/EV by converting the distributional mismatch into an average energy offset per EV.

\begin{figure}[H]
	\vspace{-4mm}
	\subfigcapskip=-2pt
	\centering
	\subfigure[Total Costs]{\includegraphics[width=0.49\columnwidth]{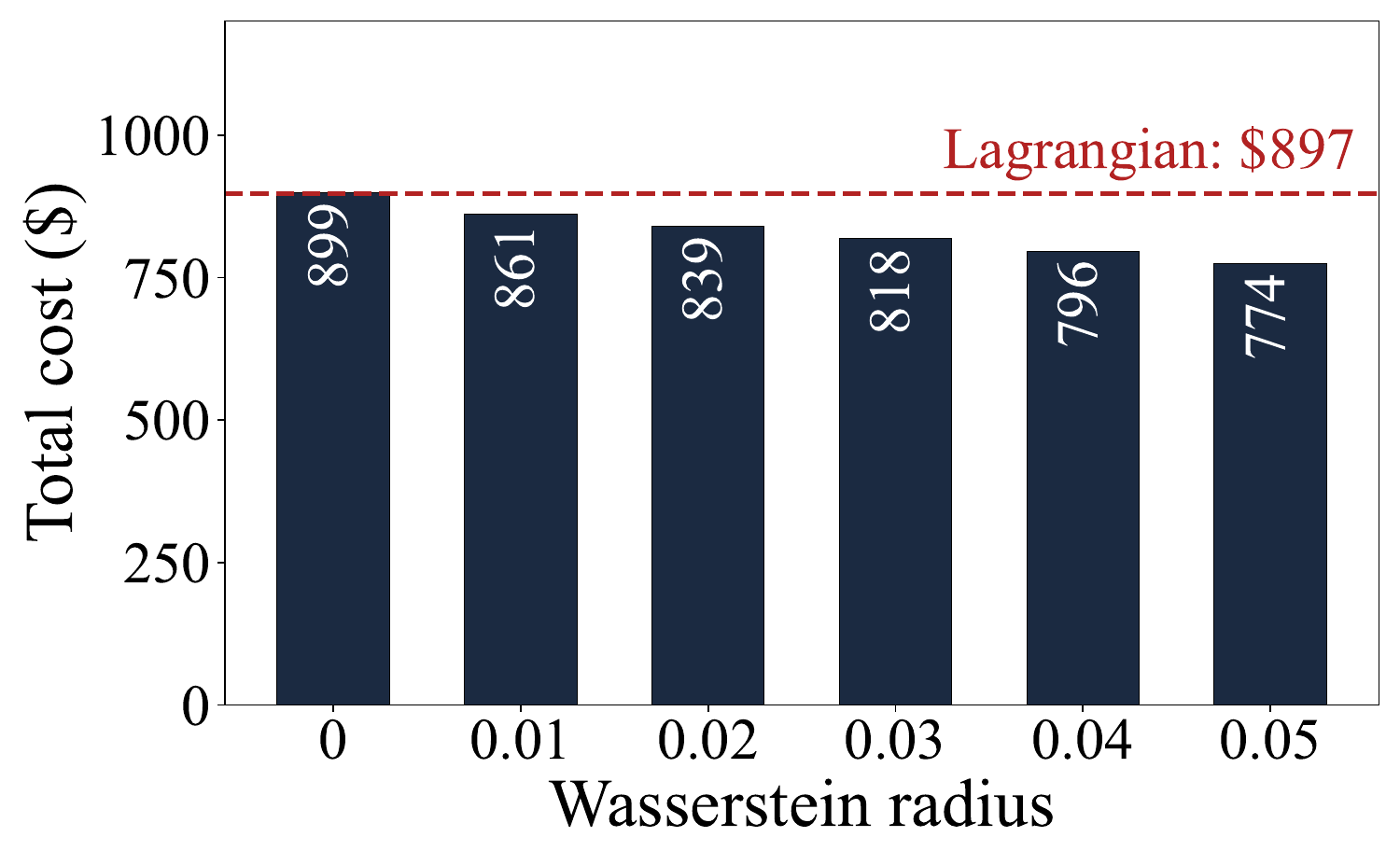}}
	\subfigure[Violation of Cyclic Constraint]{\includegraphics[width=0.49\columnwidth]{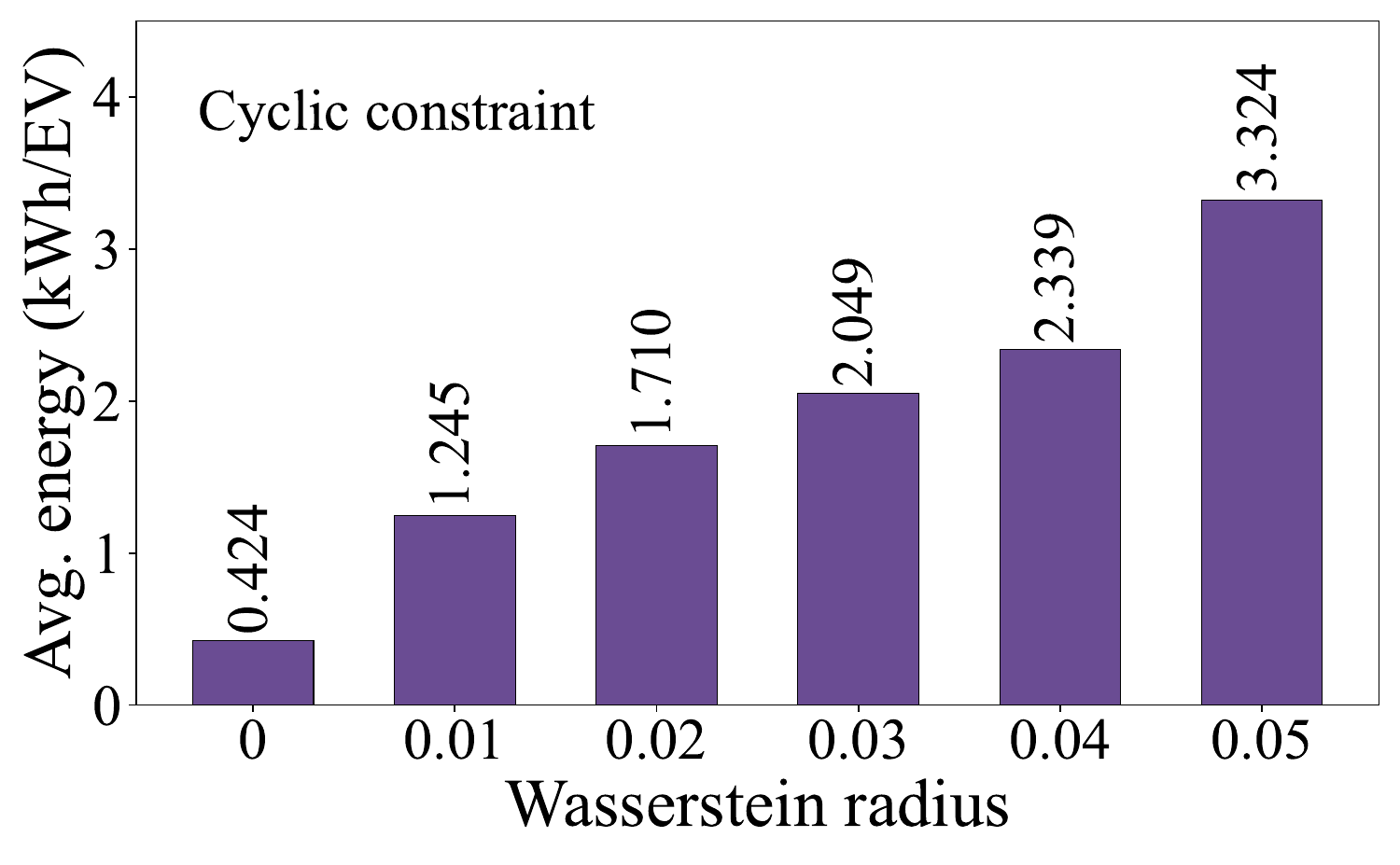}}
	\vspace{-4mm}
	\caption{Total operational costs and cyclic constraint violations under varying Wasserstein relaxation radius $\varepsilon^{\mathrm{cyc}}$ as defined in \eqref{eq:w1_ball}.}
	\label{fig_cost_EV_different_eps}
	\vspace{-4mm}
\end{figure}

Fig.~\ref{fig_cost_EV_different_eps} illustrates the impact of $\varepsilon^{\mathrm{cyc}}$ on system-level performance. As $\varepsilon^{\mathrm{cyc}}$ increases, the total operational cost decreases monotonically while the cyclic deviation grows. This confirms that relaxing near-cyclicity unlocks additional intertemporal flexibility at the expense of a larger residual end-of-horizon energy offset.

%
%

\subsection{Case II: Coordination of TCL Population}
\subsubsection{Default Parameter Settings}
\label{subsubsec:tcl_params}

We next consider the coordination of a TCL population with moderate microscopic parameter dispersion. Consistent with the homogeneity / representative-population assumption in Section~\ref{sec:modeling}, the macroscopic scheduler still uses the representative single-population Eulerian model, while parameter dispersion is reflected in the Monte-Carlo device-level execution. Unless otherwise stated, the default parameters are $N=1000$, $T=24$~h, $\Delta t=1$~min, and $K=200$. Unlike EVs (for which $f(x)=0$), TCLs exhibit thermal leakage and therefore a nonzero natural drift. The comfort band is set to $\theta^{\min}=22^\circ\mathrm{C}$ and $\theta^{\max}=26^\circ\mathrm{C}$, with ambient temperature $\theta^{\mathrm{amb}}=28^\circ\mathrm{C}$ and leakage coefficient $\alpha=0.04$. To reflect moderate within-population dispersion, the effective energy capacity is sampled as $E^{\mathrm{cap}}_i\sim\mathcal{N}(20,2^2)$~kWh, the initial normalized indoor temperature is sampled as $x_i(0)\sim\mathcal{N}(0.4,0.1^2)$, and the diffusion coefficient is set to $D=10^{-4}$.

\subsubsection{Evaluation under Varying Population Scales}


We evaluate the proposed method under varying TCL population scales and compare it with the Lagrangian benchmark, following the same settings as in the EV case. The base load profile is derived from the original curve in Fig.~\ref{fig_Parameters}, scaled proportionally to both the individual TCL capacity and the total population size. The renewable-generation profile and PCC-exchange limits are scaled in the same way so that different population sizes remain under comparable per-device operating conditions.

\begin{figure}[H]
	\vspace{-6mm}
	\subfigcapskip=-2pt
	\centering
	\subfigure[Total Costs]{\includegraphics[width=0.48\columnwidth]{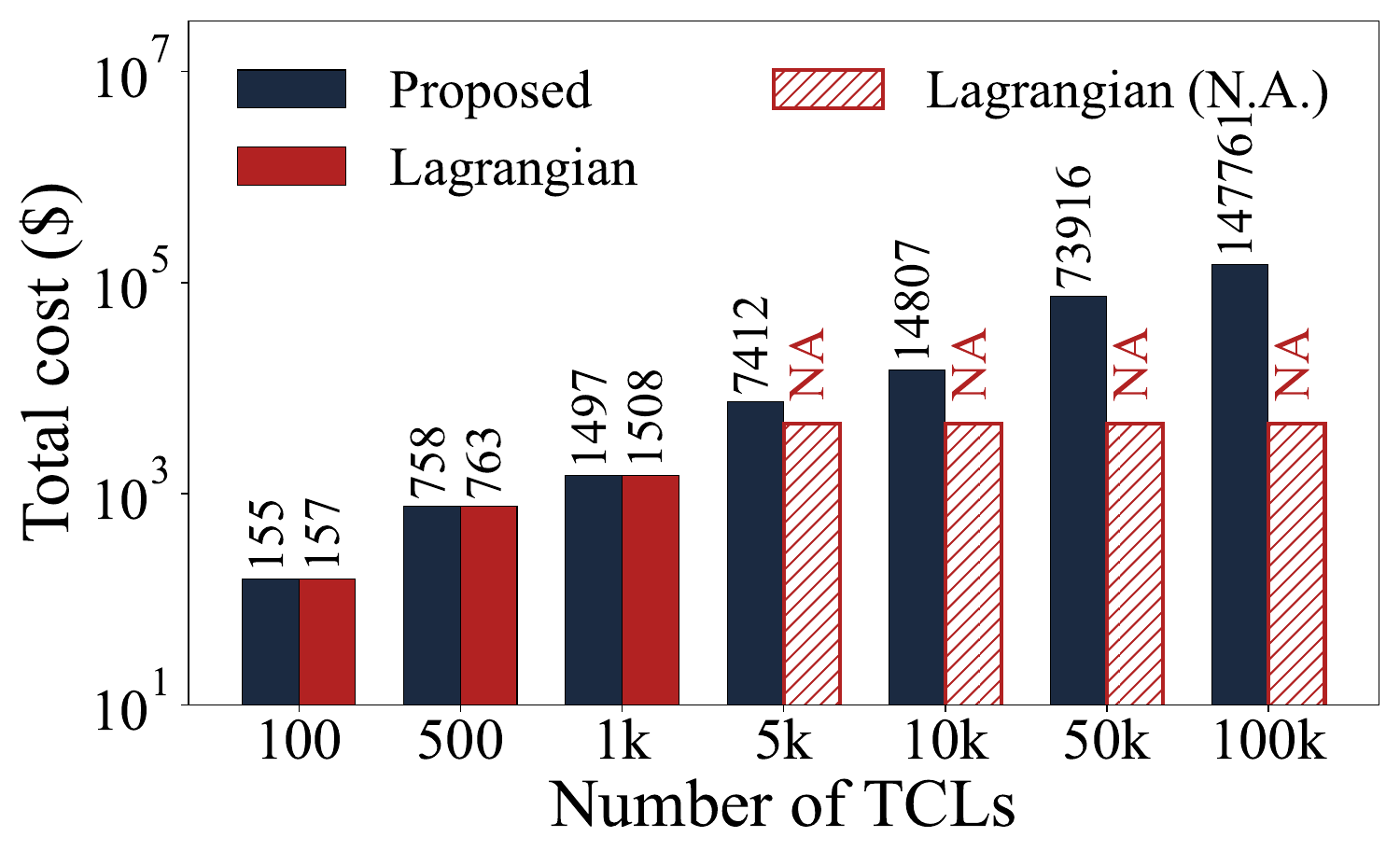}}
	\subfigure[Solving Times]{\includegraphics[width=0.5\columnwidth]{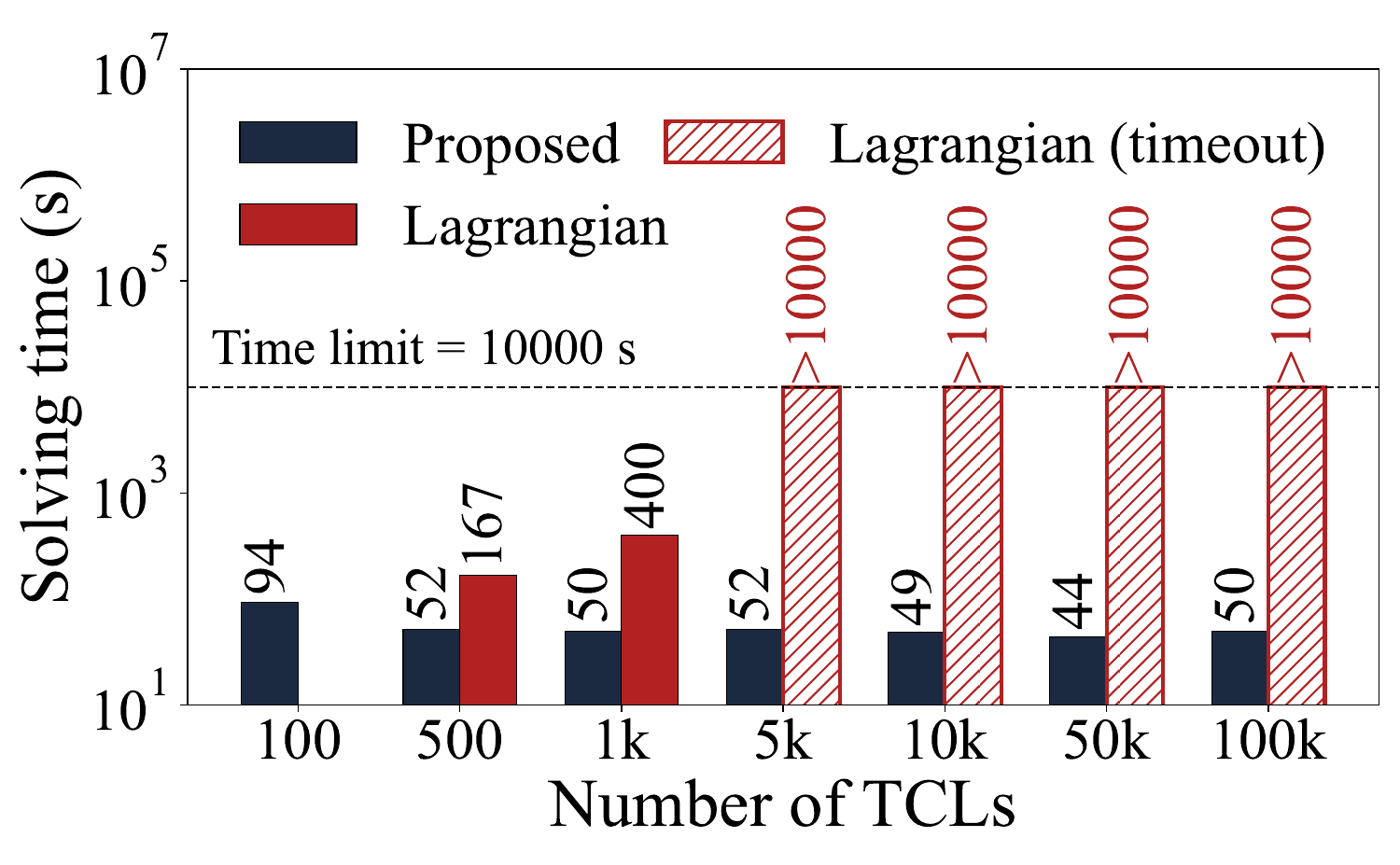}}
	\vspace{-4mm}
	\caption{Comparison of total costs and solving times between the proposed Eulerian method and the Lagrangian benchmark across varying TCL population scales. Similarly, the Lagrangian benchmark results for populations of 5,000 TCLs and above are omitted, as the solver failed to find the optima within the 10,000 s time limit.}
	\label{fig_cost_TCL_different_N}
	\vspace{-4mm}
\end{figure}

Fig.~\ref{fig_cost_TCL_different_N} summarizes the total costs and solving times across different population sizes. When the Lagrangian benchmark can be solved within the time limit (e.g., up to $N=5{,}000$), the proposed method achieves comparable total costs with substantially lower computation time, despite the stricter per-device cyclicity retained by the benchmark. As the population size increases, the benchmark quickly becomes intractable and fails to return optima within the $10{,}000$~s limit for $N\ge 5{,}000$, whereas the proposed method remains efficient across all tested sizes. This again reflects that the macroscopic LP dimension is set by the discretization rather than by the number of TCLs.
The TCL case is particularly informative because it includes nonzero natural drift and energy leakage. The close economic agreement therefore indicates that the proposed Eulerian formulation can capture not only lossless storage shifting, but also dissipative thermal flexibility with comparable scalability benefits.

\begin{figure}[H]
	\vspace{-4mm}
	\subfigcapskip=-2pt
	\centering
	\subfigure[State-Bound Constraint]{\includegraphics[width=0.49\columnwidth]{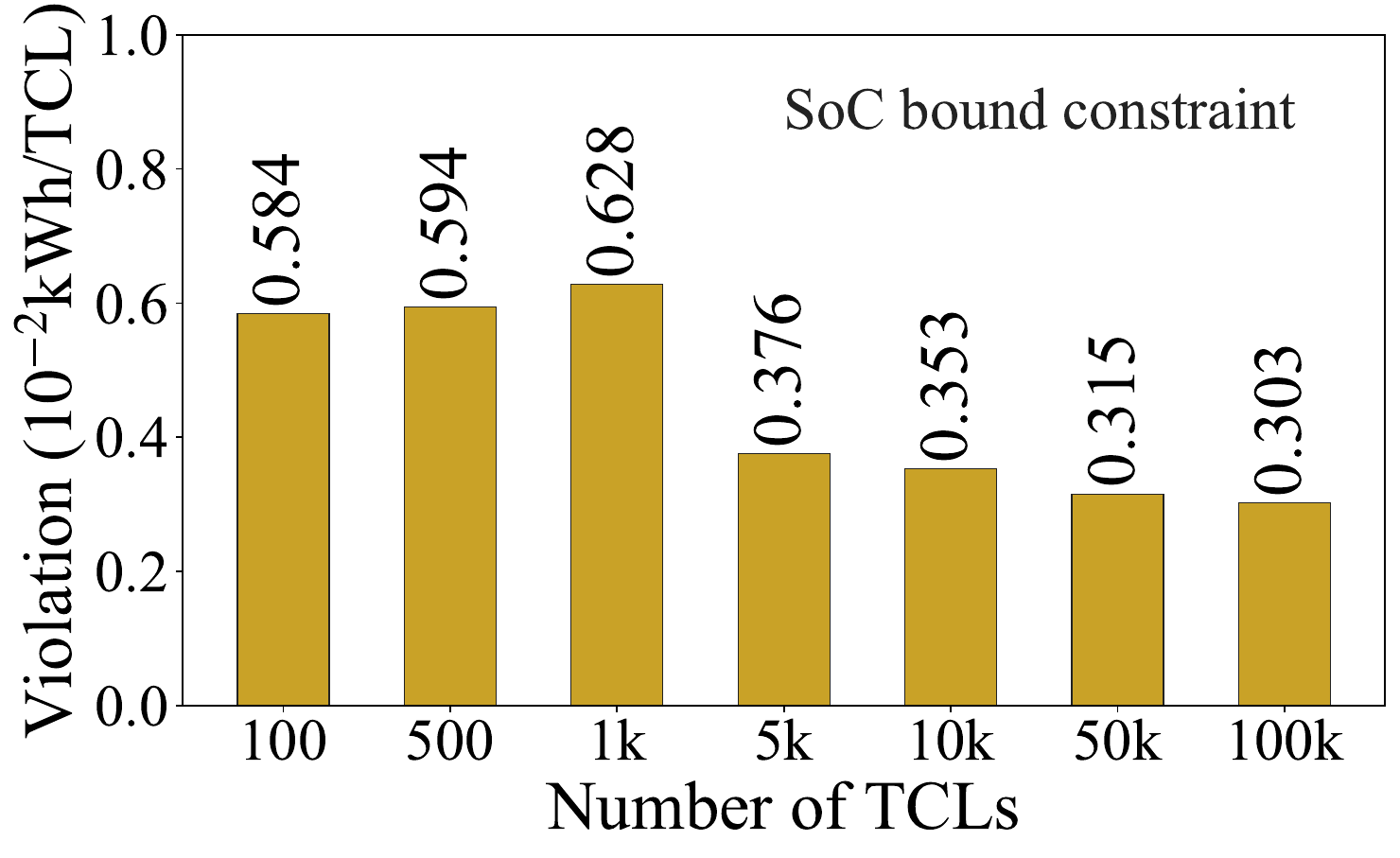}}
	\subfigure[Cyclic Constraint]{\includegraphics[width=0.49\columnwidth]{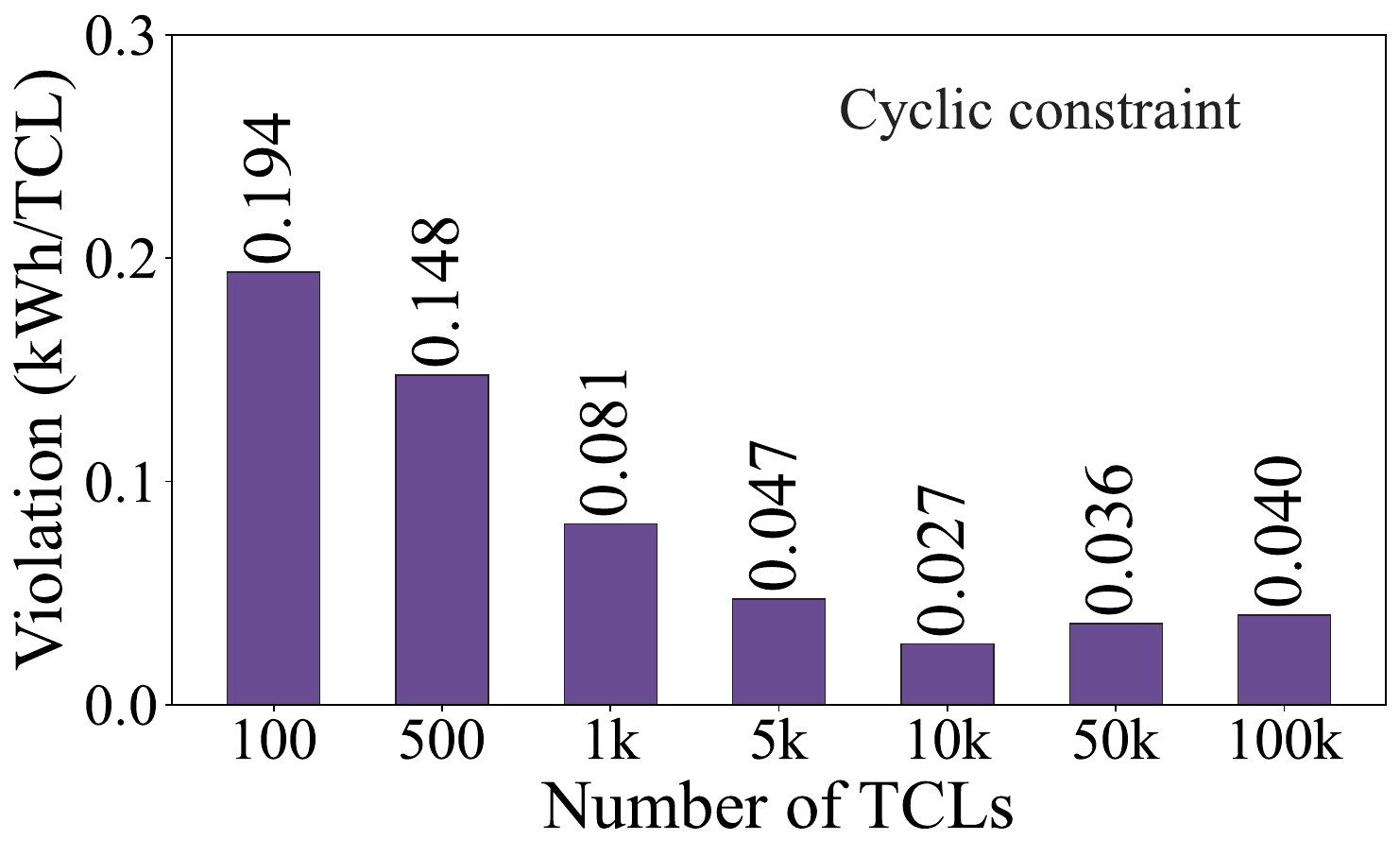}}
	\vspace{-4mm}
	\caption{Average violations of state-bound and cyclic constraints per TCL incurred by the proposed Eulerian method across varying population scales.}
	\label{fig_vio_TCL_different_N}
	\vspace{-2mm}
\end{figure}

Fig.~\ref{fig_vio_TCL_different_N} reports the state-bound and cyclic deviations per TCL under Monte-Carlo decentralized execution. Both metrics remain small across all scales, and the cyclic deviation decreases as $N$ increases, supporting the implementability of the proposed method for large-scale TCL populations.

\begin{figure}[H]
	\vspace{-2mm}
	\centering	{\includegraphics[width=1\columnwidth]{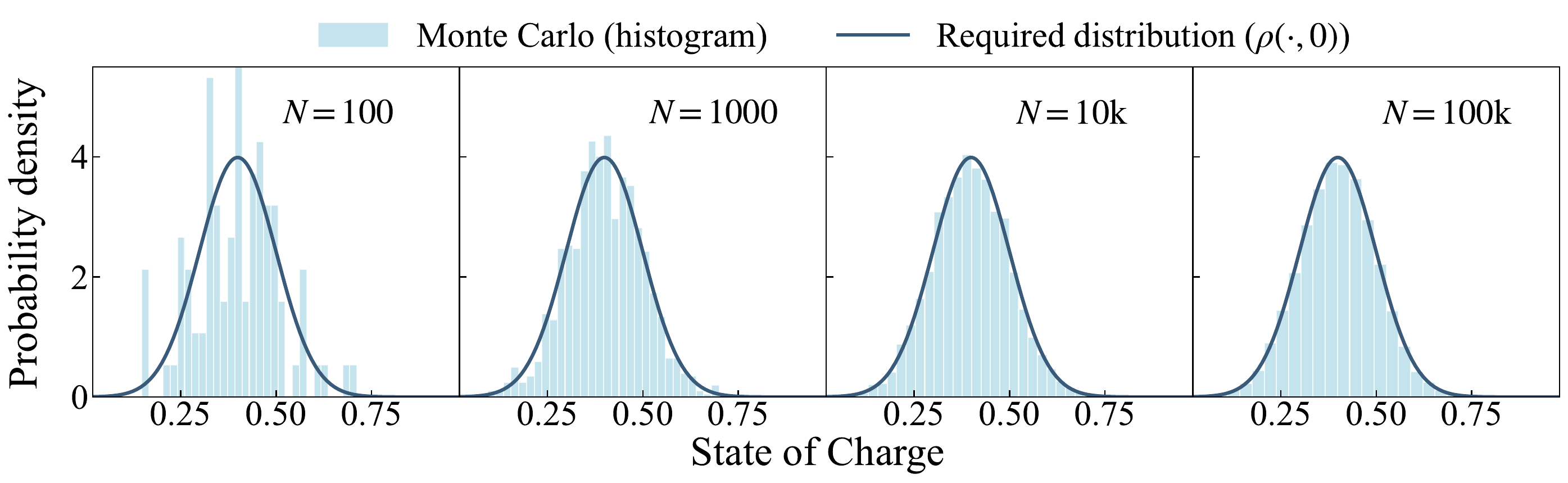}}
	\vspace{-6mm}
	\caption{The target terminal state distribution required by the cyclic constraint (blue curve) and the realized distribution obtained via Monte-Carlo simulations (density histogram) across varying TCL population scales.}
	\vspace{-2mm}
	\label{fig_cyclic_different_N_TCL}
\end{figure}

Fig.~\ref{fig_cyclic_different_N_TCL} validates the distribution-level consistency between the macroscopic schedule and the microscopic execution. As $N$ increases from $100$ to $100{,}000$, the realized terminal distributions increasingly align with the cyclicity requirement, consistent with the decreasing cyclic deviation in Fig.~\ref{fig_vio_TCL_different_N}.

\subsubsection{Impact of the Wasserstein Relaxation Radius}

\begin{figure}[H]
	\vspace{-6mm}
	\subfigcapskip=-2pt
	\centering
	\subfigure[Total Costs]{\includegraphics[width=0.49\columnwidth]{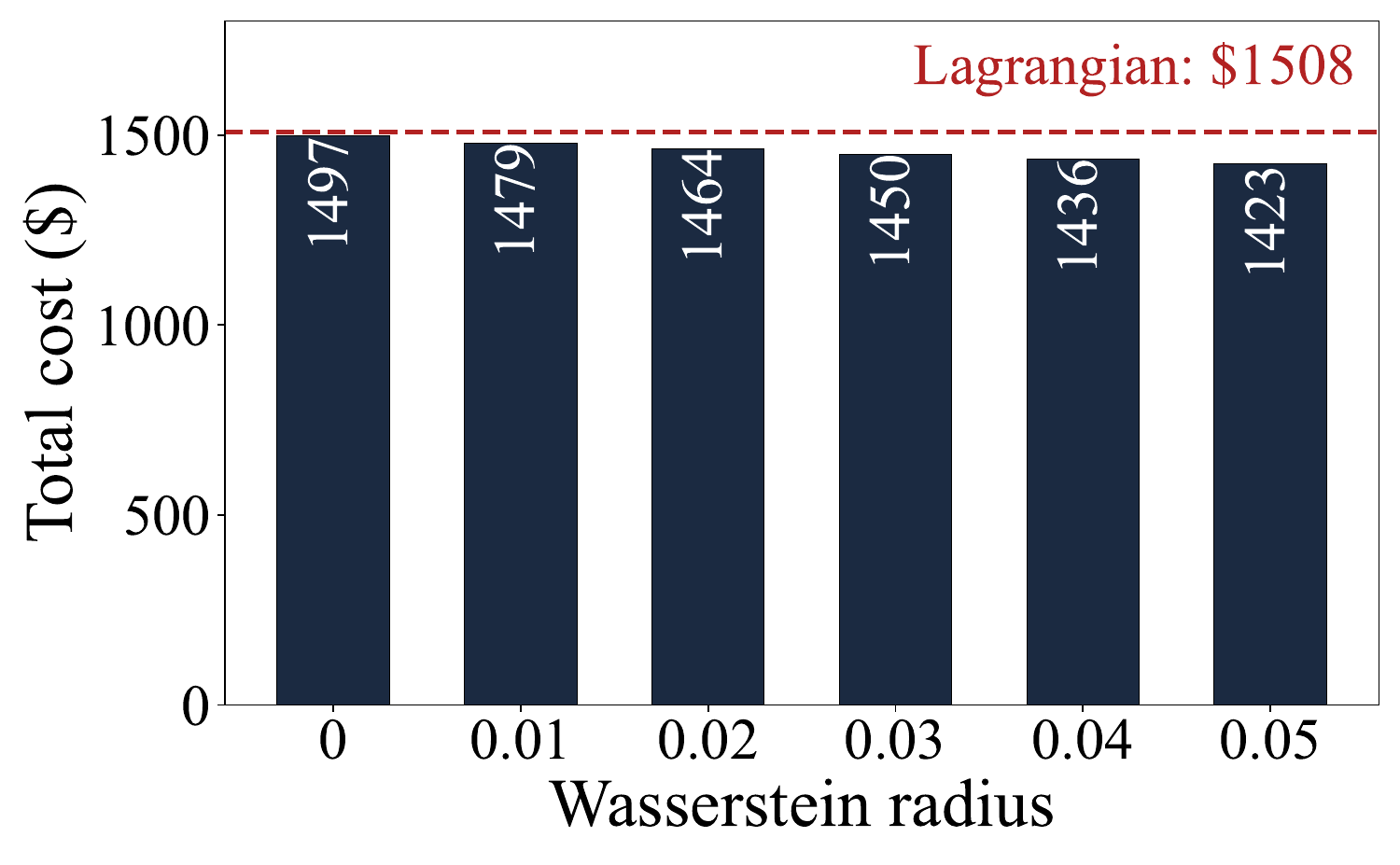}}
	\subfigure[Violation of Cyclic Constraint]{\includegraphics[width=0.49\columnwidth]{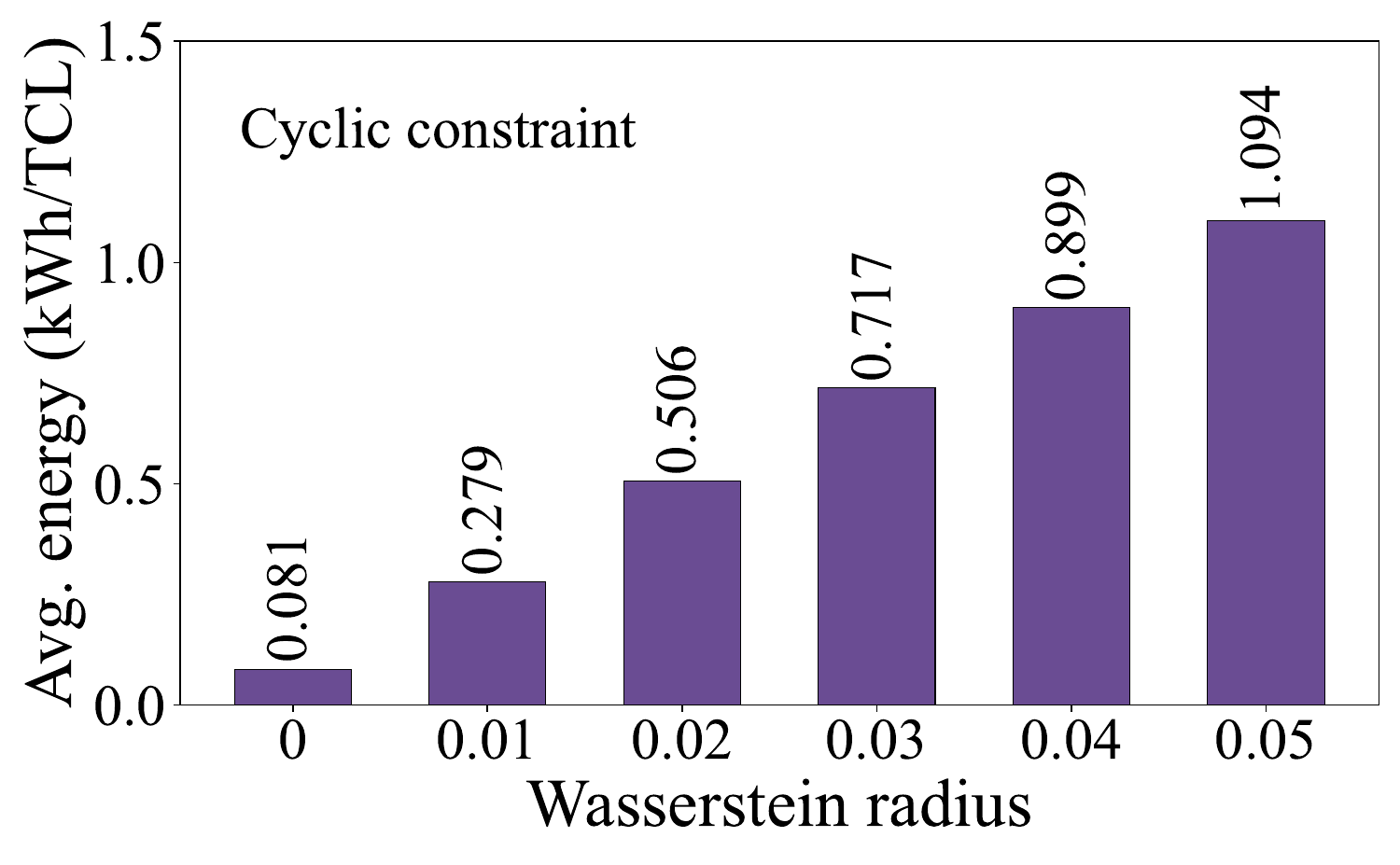}}
	\vspace{-4mm}
	\caption{Total operational costs and cyclic constraint violations under varying Wasserstein relaxation radius $\varepsilon^{\mathrm{cyc}}$ as defined in \eqref{eq:w1_ball}.}
	\label{fig_cost_TCL_different_eps}
	\vspace{-2mm}
\end{figure}

We next study the impact of the Wasserstein relaxation radius $\varepsilon^{\mathrm{cyc}}$ on the TCL population case. Fig.~\ref{fig_cost_TCL_different_eps} shows that relaxing near-cyclicity monotonically reduces the total operating cost while increasing the realized cyclic deviation, yielding the same economics-cyclicity trade-off observed in the EV case. The TCL results also provide a physical interpretation of this relaxation: allowing a moderate terminal-state bias lets the population end the horizon in thermally favorable states, which reduces leakage-related cost while remaining within the relaxed cyclicity envelope.

%

\section{Conclusion}
\label{sec:conclusion}

This paper develops a two-layer Eulerian coordination framework for large-scale storage-like DER populations. At the macroscopic layer, the aggregator optimizes the evolution of the population state density through an Eulerian model, whose size is governed by the discretization resolution rather than by the population size. To overcome the inherent bilinear non-convexity, finite-volume discretization and flux lifting are combined to reformulate the discretized scheduling problem as a tractable sparse LP. A Wasserstein-based near-cyclicity constraint is further introduced to relax rigid terminal recovery and provide the aggregator with a tunable trade-off between end-of-horizon recovery and economic benefit. At the microscopic layer, individual devices autonomously recover their power setpoints from the broadcast state-dependent signal and their local states, while a histogram-based data-mixing interface reduces direct aggregator access to raw individual states. Case studies on EV and TCL populations demonstrate strong scalability and good practical feasibility, with small economic gaps relative to the device-level reference.

The current study focuses on single or representative populations and does not explicitly model detailed network constraints or discrete ON/OFF resources. Future work will extend the framework to multi-population coordination, network-aware operation, and richer heterogeneous device models.

\appendices
\section{}
\label{app:fvm_ops}

Let $\Delta x = 1/K$ and $\mu \triangleq D \Delta t / \Delta x^2$. The discrete operators used in Section~\ref{sec:solver} are defined as follows.

\subsection{\texorpdfstring{Divergence Matrix $\mathbf{M}$}{Divergence Matrix M}}

Matrix $\mathbf{M} \in \mathbb{R}^{K \times (K+1)}$ maps interface fluxes to cell-centered density changes with entries:
\begin{align}
	(\mathbf{M})_{k,l} = \frac{\Delta t}{\Delta x} \times 
	\begin{cases} 
		-1, & l = k, \\ 
		1, & l = k+1, \\ 
		0, & \text{otherwise}.
	\end{cases}
\end{align}

\subsection{\texorpdfstring{Diffusion Matrix $\mathbf{L}$}{Diffusion Matrix L}}
Matrix $\mathbf{L} \in \mathbb{R}^{K \times K}$ satisfies $-\mathbf{L}\boldsymbol{\rho}_{t+1} = \mathbf{M}\boldsymbol{\Phi}^{\mathrm{dif}}_{t+1}$ under boundary condition \eqref{eq:bc_compact}:
\begin{itemize}
	\item Internal rows ($k=2, \dots, K-1$): the entry is $(L)_{k,k} = -2\mu$ and $(L)_{k,k \pm 1} = \mu$.
	\item Boundary rows: the entry is $(L)_{1,1} = -\mu, (L)_{1,2} = \mu$ and $(L)_{K,K} = -\mu, (L)_{K,K-1} = \mu$.
\end{itemize}

\subsection{\texorpdfstring{Averaging Matrix $\mathbf{A}^{\mathrm{avg}}$}{Averaging Matrix Aavg}}
Matrix $\mathbf{A}^{\mathrm{avg}} \in \mathbb{R}^{K \times (K+1)}$ maps interface fluxes to cell centers with entries:
\begin{align}
	(\mathbf{A}^{\mathrm{avg}})_{k,l} = 
	\begin{cases} 
		0.5, & l = k \text{ or } l = k+1, \\ 
		0, & \text{otherwise}.
	\end{cases}
\end{align}

\footnotesize
\bibliographystyle{ieeetr}
\bibliography{ref}

\end{document}